\documentclass[a4paper,fleqn,usenatbib]{mnras}

\usepackage[T1]{fontenc}
\usepackage{ae,aecompl}
\usepackage{array}

\usepackage{graphicx}	% Including figure files
\usepackage{amsmath}	% Advanced maths commands
\usepackage{amssymb}	% Extra maths symbols

\title[Magnetic fields in filaments (I)]{Magnetic fields in star forming systems (I): Idealized synthetic signatures of dust polarization and Zeeman splitting in filaments}

\author[Stefan Reissl]{
Stefan Reissl,$^{1}$\thanks{E-mail: reissl@uni-heidelberg.de}
Amelia M. Stutz,$^{2,3}$
Robert Brauer,$^{4}$
Eric W. Pellegrini,$^{1}$
\newauthor
\qquad\qquad\qquad\qquad Dominik R.G. Schleicher,$^{2}$
and Ralf Klessen$^{1,5}$
\\
$^{1}$Heidelberg University, Center for Astronomy, Institute of Theoretical Astrophysics, Albert-Ueberle-Str. 2, 69120 Heidelberg, Germany\\
$^{2}$Departamento de Astronom\`ia, Facultad Ciencias F\`isicas y Matem$\acute{a}$ticas, Universidad de Concepci$\acute{o}$n, \\\qquad Av. Esteban Iturra s/n Barrio Universitario, Casilla 160-C, Concepci$\acute{o}$n 4030000, Chile \\
$^{3}$Max-Planck-Institute for Astronomy, K\"onigstuhl 17, 69117 Heidelberg, Germany\\
$^{4}$CEA Saclay - DRF/IRFU/SAp, Orme des Merisiers, B$\hat{a}$t 709, 91191 Gif sur Yvette\\
$^{5}$Universit{\"a}t Heidelberg, Interdisziplin{\"a}res Zentrum f{\"u}r Wissenschaftliches Rechnen, Im Neuenheimer Feld 205, 69120 Heidelberg, Germany
}

\date{Accepted XXX. Received YYY; in original form ZZZ}

\pubyear{2018}

\begin{document}
\label{firstpage}
\pagerange{\pageref{firstpage}--\pageref{lastpage}}
\maketitle

\begin{abstract}
We use the POLARIS radiative transport code to generate predictions of the two main observables directly sensitive to the magnetic field morphology and strength in filaments: dust polarization and gas Zeeman line splitting.  We simulate generic gas filaments with power-law density profiles assuming two density-field strength dependencies, six different filament inclinations, and nine distinct magnetic field morphologies, including helical, toroidal, and warped magnetic field geometries. We present idealized spatially resolved dust polarization and Zeeman-derived field strengths and directions maps.  Under the assumption that dust grains are aligned by radiative torques (RATs), dust polarization traces the projected plane-of-the-sky magnetic field morphology. Zeeman line splitting delivers simultaneously the intensity-weighted line-of-sight field strength and direction. We show that linear dust polarization alone is unable to uniquely constrain the 3D field morphology.  We demonstrate that these ambiguities are ameliorated or resolved with the addition of the Zeeman directional information.  Thus, observations of both the dust polarization and Zeeman splitting together provide the most promising means for obtaining constraints of the 3D magnetic field configuration.  We find that the Zeeman-derived field strengths are at least a factor of a few below the input field strengths due to line-of-sight averaging through the filament density gradient.  Future observations of both dust polarization and Zeeman splitting are essential for gaining insights into the role of magnetic fields in star and cluster forming filaments.
\end{abstract}


%%%%%%%%%%%%%%%%%%%%%%%%%%%%%%%%%%%%%%%%%%%%%%%%%%

%%%%%%%%%%%%%%%%% BODY OF PAPER %%%%%%%%%%%%%%%%%%

\section{Introduction}
The study of gas filaments has virtually exploded in the field of star
and cluster formation in the last $\sim$ half decade.  This explosion
has been predominately driven by data from the {\it Herschel} satellite, which showed with
more undeniable empirical clarity than previously available that main
structural components of star-forming molecular clouds are filaments
\citep[e.g.,][]{Molinari2010,Andre2010,Rathborne2011,Arzoumanian2011,Stutz2015,Stutz2016}. At
the same time, driven by the observational basis provided most
recently by the Planck mission \citep[][]{Tauber2010,Planck2011}, the
study of the observational signatures of the magnetic field with
polarization is receiving renewed and heated interest in these
filamentary star forming regions.  While the Planck data have
comparatively low angular resolution, their contribution to the
discussion of filament structure through dust polarization
observations is undeniable \citep[][]{Planck2015XX,Planck2016XXXV}.%\citep[][]{Li2015}
ALMA mosaic observations will be able to probe more distant and massive filaments throughout the Galaxy using both (sub)millimeter dust polarization and Zeeman line splitting observation . Already a variety of single dish results have
paved the way for such observations
\citep[e.g.,][]{Matthews2001,Matthews2002,Bertrang2014,Pillai2015,Pattle2017}.

On the theoretical side, a wealth of studies going back to
\cite{Chandrasekhar1953} have explored the possible roles of magnetic
fields in combination with turbulence in the ISM and specifically in
filaments
\citep[e.g.][]{Nagasawa1987,Fiege2000a,Tomisaka2014,Toci2015a,Schleicher2018}. The
advent of improved turbulence simulations
\citep[e.g.,][]{chen2015,Walch2015,Seifried2017,ibanez2017,ntormousi2017,iffrig2017,inoue2017}
with the inclusion, if in a simplified fashion, of magnetic fields in
the form of an additional source of pressure in the gas have further
driven forward the study of the possible role of magnetic fields in
such structures in the ISM.  In these works, various scenarios have
been proposed, as well as various field configurations, both on the
basis of direct observations (which have in the past been
comparatively limited) and on the basis of what is "observed" in the
numerical simulations \citep[see][for review]{Klessen2016}.

\begin{figure*}
\begin{center}
        \begin{minipage}[c]{1.06\linewidth}
                        \begin{center}
                                \includegraphics[width=0.35\textwidth]{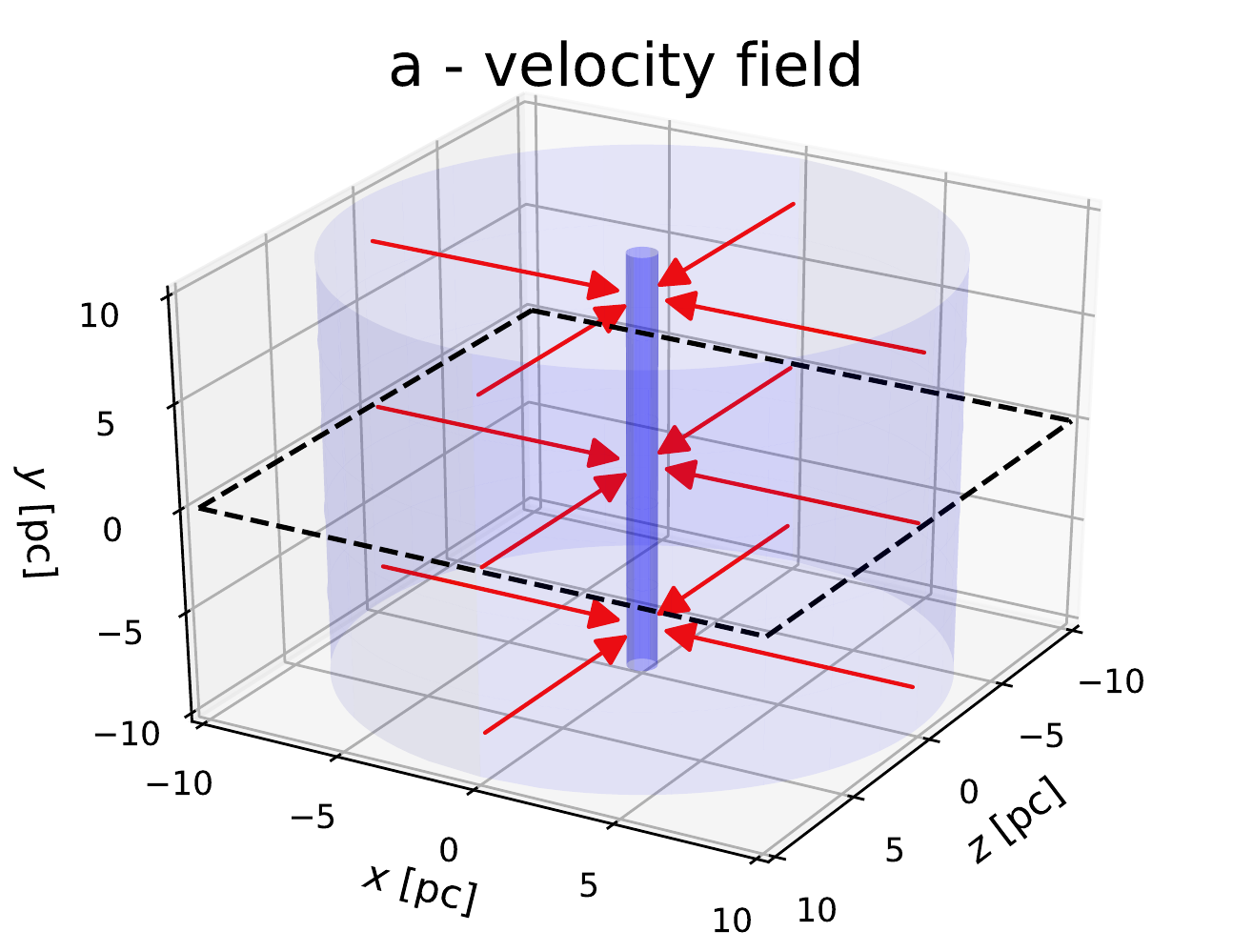}
                                \hspace{-7.38 mm}
                                \includegraphics[width=0.35\textwidth]{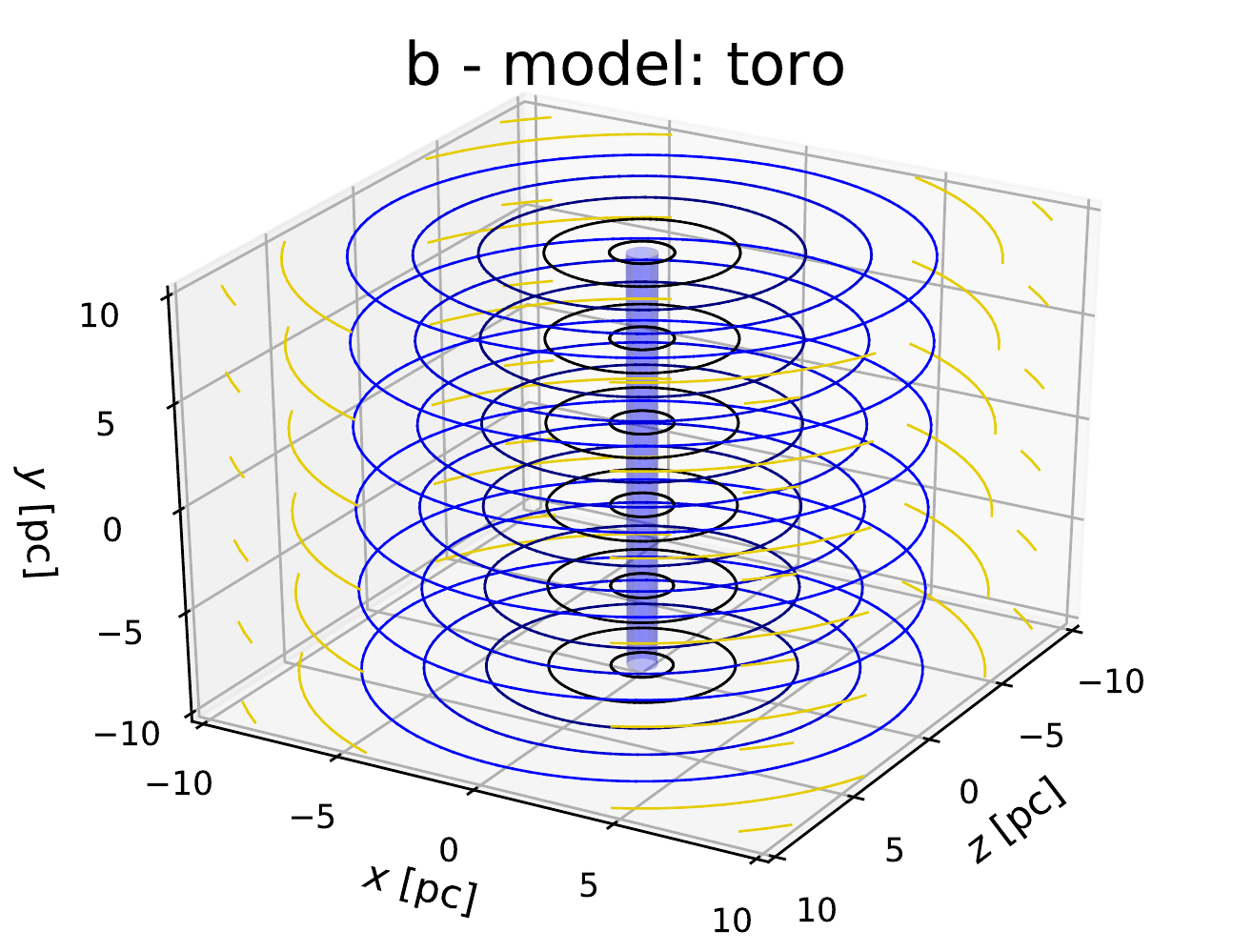}
                                \hspace{-7.38 mm}
                                \includegraphics[width=0.35\textwidth]{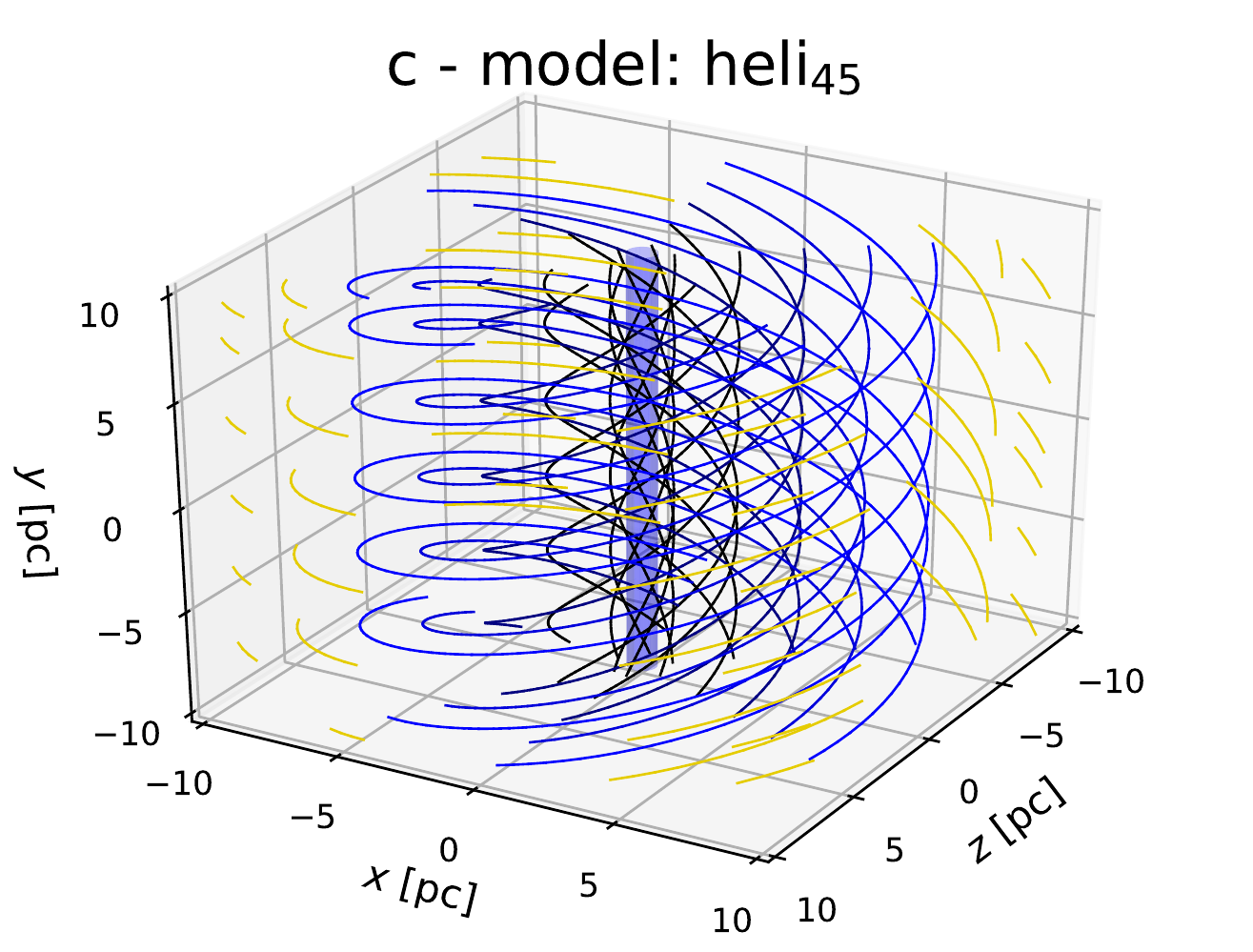}
                        \end{center}
        \end{minipage}
        
        \vspace{1 cm}
       
       \begin{minipage}[c]{1.06\linewidth}
                        \begin{center}
				\includegraphics[width=0.35\textwidth]{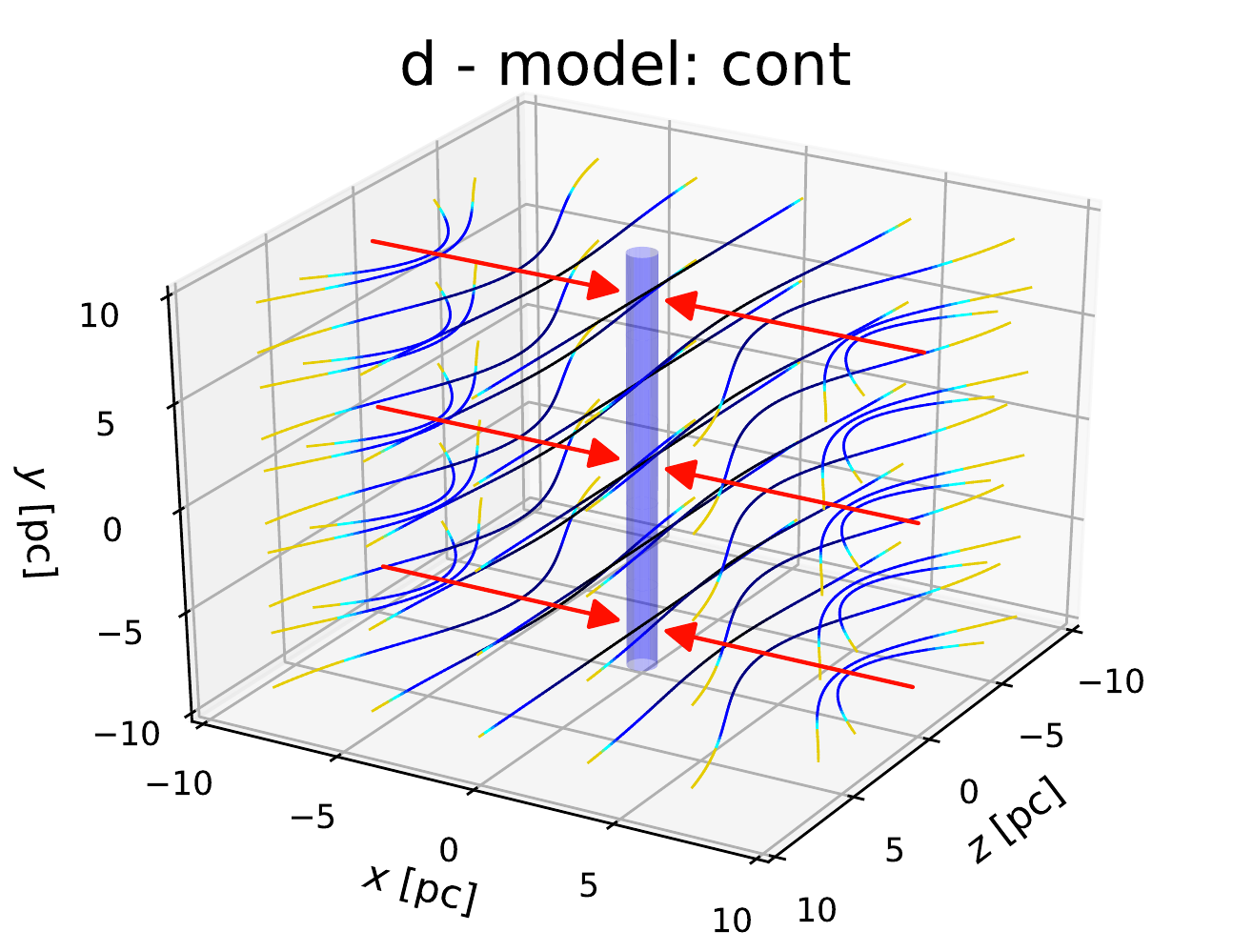}
				\hspace{-7.38 mm}
                                \includegraphics[width=0.35\textwidth]{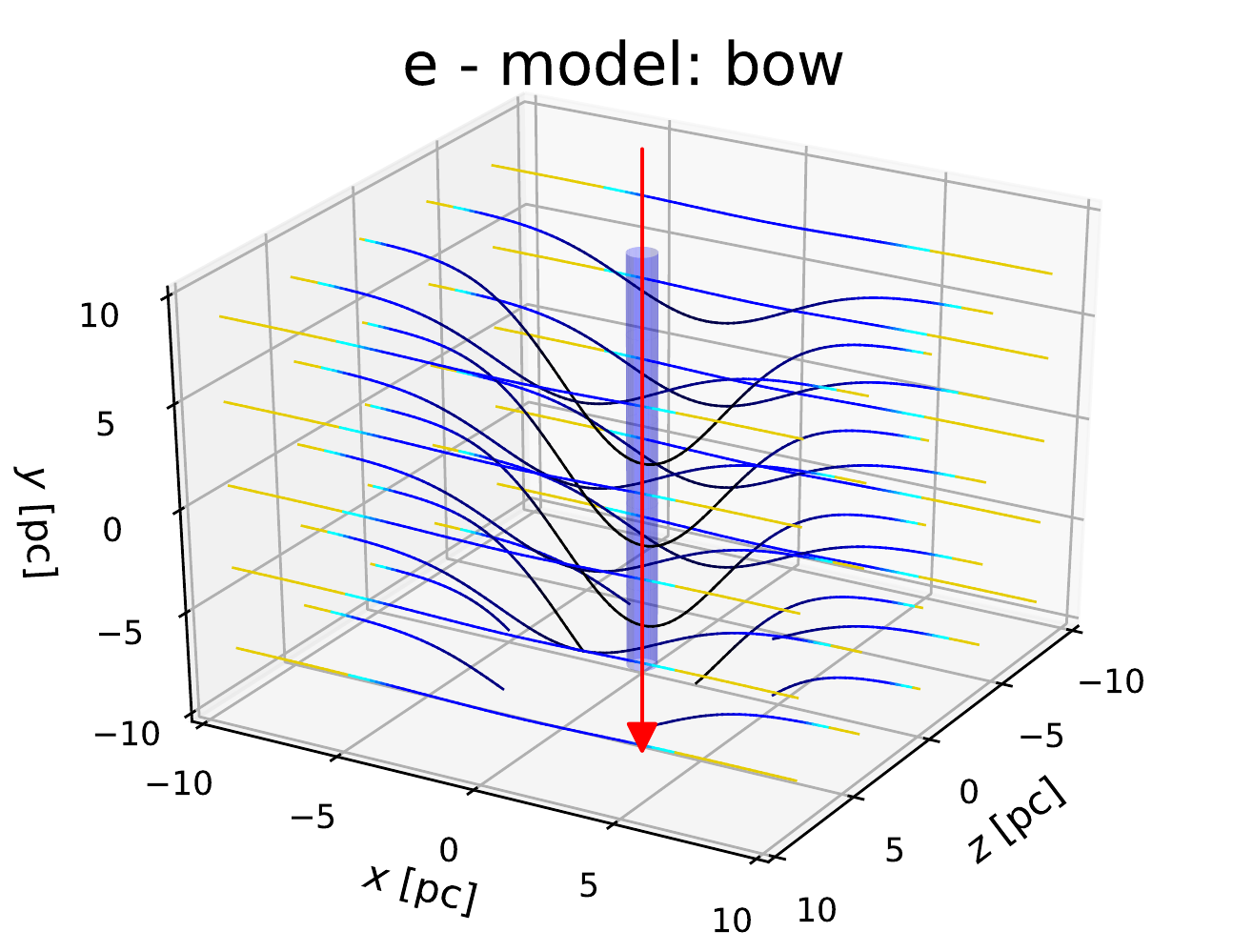}
                                \hspace{-7.38 mm}
                                \includegraphics[width=0.35\textwidth]{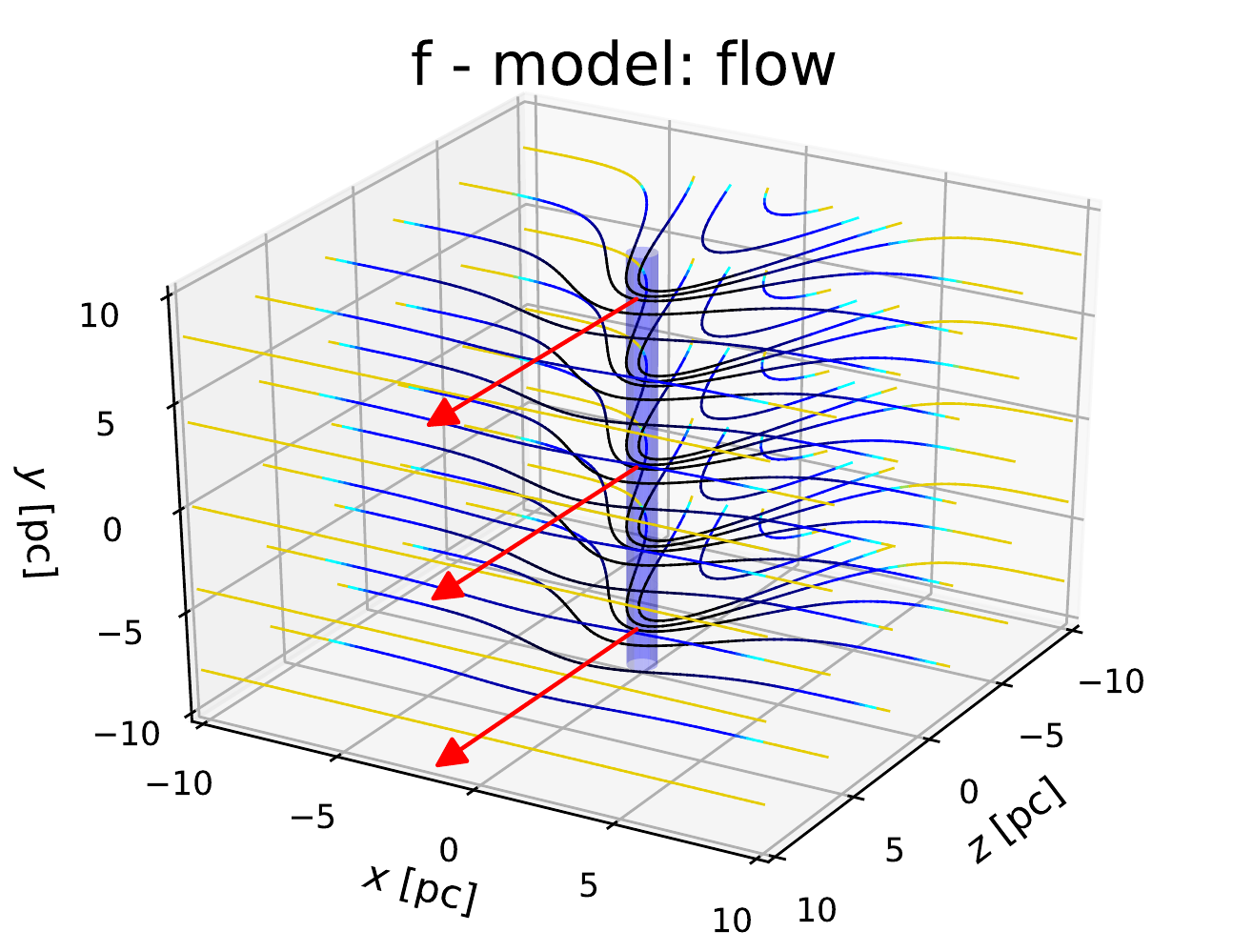}
                        \end{center}
       \end{minipage}       
       \begin{minipage}[c]{1.0\linewidth}
                        \begin{center}
				\includegraphics[width=0.66\textwidth]{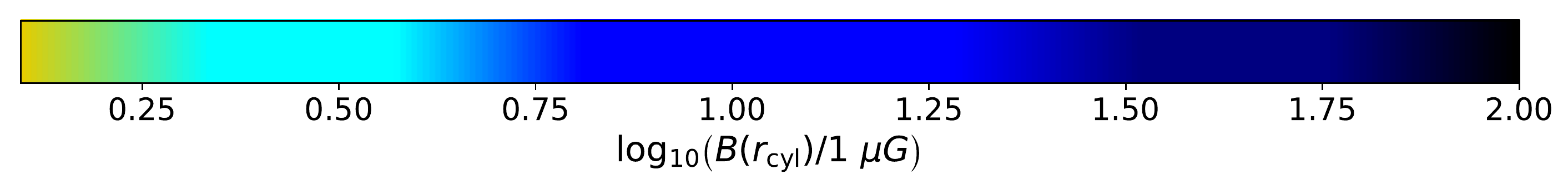}
	                 \end{center}
                       \end{minipage} \end{center} 

\caption{Panel a: Model cube with a side length of $\pm
  10\ \rm{pc}$. The isosurface represents the gas number density
  $n_{\rm gas}(r_{\rm cyl})$ (see Eq. \ref{eq:DensityDistribution}) at
  a distance of $r_{\rm cyl}=0.25$ and $r_{\rm cyl}=1$, respectively,
  from the symmetry axis. The red arrows indicate the orientation of
  the velocity field of the slightly collapsing filament, while the
  black dashed lines indicates the observer plane along which all
  synthetic observations are derived. The direction of observation is
  by default along the z-axis while all rotations are performed around
  the x-axis. Panel b: The magnetic field model $'\rm{toro}'$ modeled
  with Eq. \ref{eq:MagToro}. Panel c: Representation of the class of
  helical magnetic field model $'\rm{heli}_{\rm \alpha}'$ modeled with
  Eq. \ref{eq:MagHeli} shown for a pitch angle of
  $\alpha=45^{\circ}$. Panel d: The magnetic field model $'\rm{cont}'$
  modeled with Eq. \ref{eq:MagCont}. Red arrows indicate the velocity
  components of the velocity field presented in panel a. that can drag
  the magnetic field lines. Panel e: The magnetic field model
  $'\rm{cont}'$ modeled with Eq. \ref{eq:MagBow}. The red arrow
  indicates the contraction of the filament along the symmetry
  axis. Panel f: The magnetic field model $'\rm{flow}'$ modeled with
  Eq. \ref{eq:MagFlow}. The red arrows indicate the additional
  velocity component with which the filament is drifting into the
  initially straight magnetic field morphology.}
\label{fig:Models}
\end{figure*}

The main configurations that have been proposed and observed for
magnetic field geometries can be divided into two principle categories
illustrated in Figure~\ref{fig:Models}.  First, there are those that
represent distortions of an approximately straight field-line
configuration, such as a bow-like or a gravitationally distorted
field.  Then there are those that are entirely curved and have closed
(or approximately closed and approximately divergence free) field
lines, such as a toroidal or helical field configuration wrapping
around the filament
\citep[e.g.,][]{heiles1987,uchida1991,tatematsu1993,heiles1997,Fiege2000a,Schleicher2018,tahani2018}.
We expect potentially radically different behaviors of a system in the
presence of these two different types of fields. For example, for a
helical field we might observe a magnetic pinch
\citep[so called "z-pinch"; e.g.,][]{bocchi2013,shtemler2009}
instability to develop, which may compress the filament material
\citep{Stutz2016,stutz2018}.  In the presence of a bow shaped field, the
interpretation of the observed geometry may lead to the conclusion
that a field is being distorted by the action of gravity and the
magnetic field is energetically sub-dominant.  Both are in principle
possible or at least have been previously proposed, yet the
consequences and implications for filamentary star and cluster
formation may be very different in each case.

\begin{figure*}
\begin{center}
        \begin{minipage}[c]{1.0\linewidth}
                        \begin{center}
                                \includegraphics[width=0.55\textwidth]{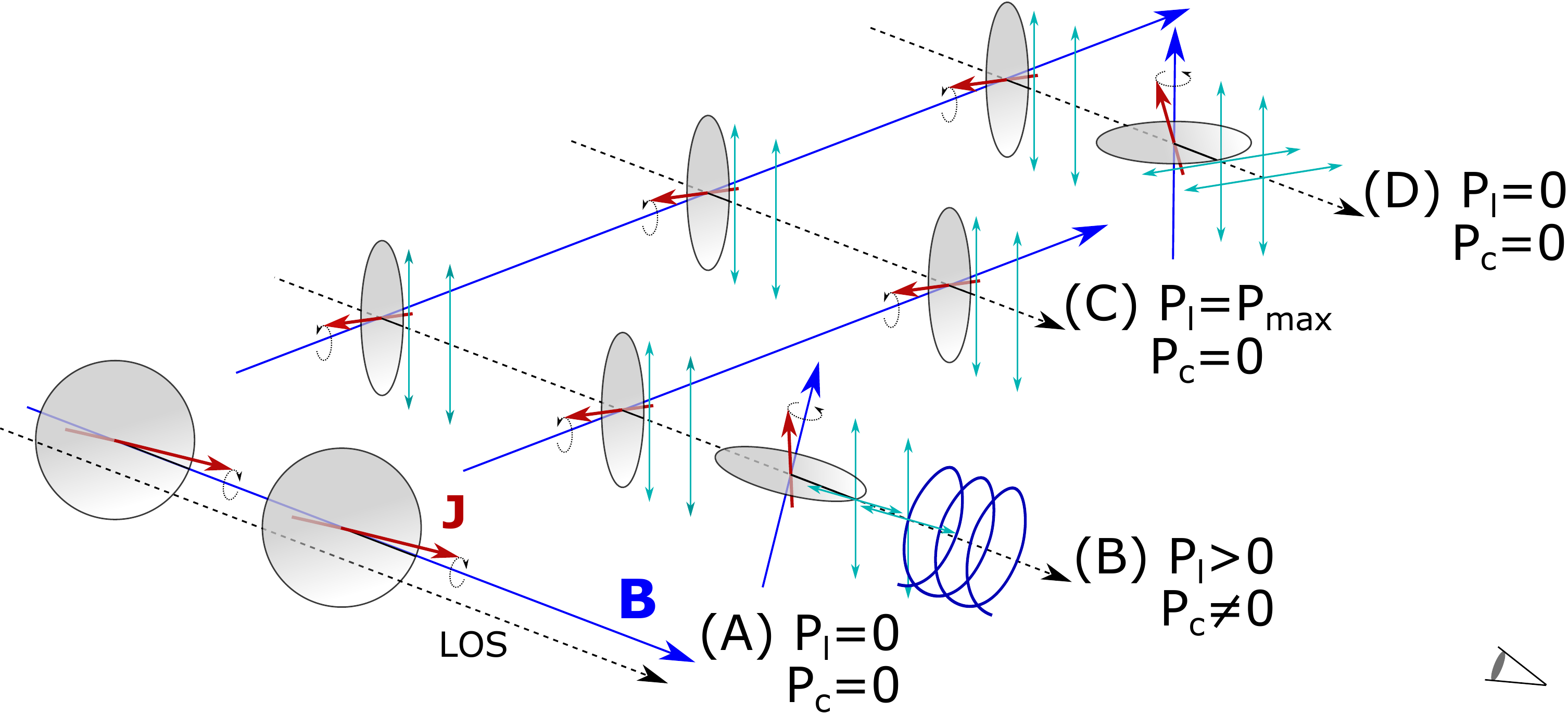}
                                \includegraphics[width=0.43\textwidth]{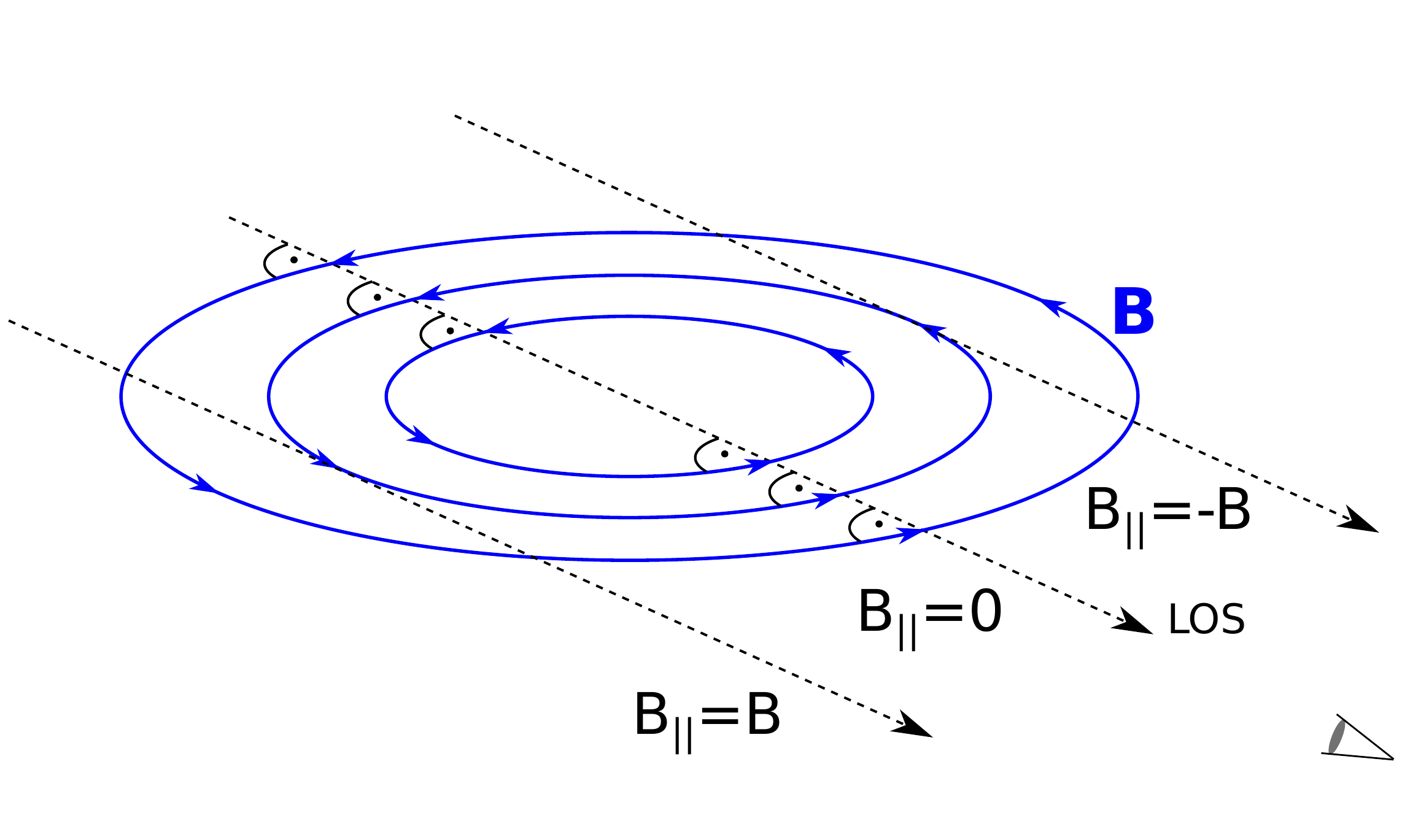}
                        \end{center}
                      \end{minipage} \end{center} 

\caption{Left panel: Sketch of aligned dust grains precessing with
  their angular momentum $\vec{J}$ (red arrows) around the magnetic
  field direction $\vec{B}$ (blue arrows) observed along the LOS
  (blacked dashed arrows). In scenario (A) $\vec{B}$ and LOS are
  parallel and the dust grains appear spherical. Hence, no linear
  $P_{\rm l}$ or $P_{\rm c}$ polarization can be observed. Scenario
  (B) shows twisted field lines. The net orientation of linear
  polarization $P_{\rm l}$ represents a superposition of all the field
  lines along the LOS and a small amount of circular polarization
  $P_{\rm c}$ accumulates. In scenario (C) are all adjacent field
  lines parallel to each other. The linear polarization $P_{\rm l}$ is
  maximal while all polarized radiation experiences the same amount of
  differential phase lag. Consequently, no circular polarization
  $P_{\rm c}$ can be built up. In case (D) two adjacent field lines
  are exactly perpendicular to each other. All contributions of
  polarized thermal dust emission cancel out. Right: Sketch of Zeeman
  observations along different LOS directions for the case of a
  toroidal magnetic field morphology $\vec{B}$. When LOS and $\vec{B}$
  are perpendicular, the magnetic field component $\vec{B}_{\rm ||}$
  can be observed. For a LOS that is parallel or anti-parallel,
  respectively, to $\rm{B}$, the component $\vec{B}_{\rm ||}$ has its
  maximum. The sign of $\vec{B}_{\rm ||}$ allows us to infer the
  parallel or anti-parallel configuration of $\rm{B}$ with respect to
  the LOS.}
\label{fig:Sketch}
\end{figure*}

In order to constrain possible underlying field configurations,
synthetic observations are an essential intermediate step in the
analysis of actual observations.  Here we focus on two observable
signatures of the magnetic field: dust polarization \citep[see][for a
  review]{Andersson2015} and Zeeman splitting of various molecular
lines \citep[e.g., ][]{Crutcher1993}.  We adopt a filament power-law
density profile consistent with observations of the Orion~A Integral
Shaped Filament (ISF; \cite{Stutz2016}) and also test other density
profiles \citep[e.g.,][]{Arzoumanian2011}. As for our adopted magnetic
field configurations we implement numerous suggestions from
observations and theoretical works, focusing on five distinct magnetic
field morphologies associated with the evolution of filaments and star
formation. We then predict their idealized observable signatures by
making use of the radiative transfer (RT) code POLARIS
\citep[][]{Reissl2016,Brauer2017}\footnote{http://www1.astrophysik.uni-kiel.de/${\sim}$polaris/}. The
POLARIS code is the first of its kind capable of simulating dust
polarization on the basis of state of the art grain alignment physics
in combination with line RT, including the Zeeman effect. From the
dust polarization signature we obtain the 2D projections of the
magnetic field in the plane of the sky ($B_{||}$), under the assumption
that the dust grains are aligned by radiative torques 
\citep[RAT;][which is the most likely cause of grain
  alignment]{Draine1996,Draine1997,Weingartner2003,Lazarian2007}. From
the Zeeman line spitting signature we obtain two pieces of
information: the line-of-sight (LOS) component of the magnetic field 
strength (estimated from simulated circular polarization profile).
This synthetic observation approach is the first to
predict the impact of the magnetic field morphology on the dust
polarization pattern and the complementary Zeeman measurements
simultaneously. The combination of these two diagnostics will prove to
be invaluable in the ultimate goal of reconstructing the 3D magnetic
field configuration in the future when both types of observations
become routine in star and cluster forming filaments.

We demonstrate in this paper that dust polarization alone provides ambiguous constraints for the magnetic field morphology. We show that Zeeman observations provide the necessary additional information to constrain the underlying 3D magnetic field morphology.

This paper is organized as follows: First, we introduce the geometry,
gas density profile, and velocity profile of our filament model in
Sect. \ref{sect:RadialProfiles}.  Then, we give a description of the
applied dust component in Sect. \ref{sect:GrainProperties} followed by
the modeling parameters of the magnetic field morphologies that we
consider in Sect. \ref{sect:FieldMorphologies}. In
Sect. \ref{sect:MolecularAbundances} we present a way of synthesizing
additional molecular abundances. The physics of dust heating and grain
alignment, as well as RT with polarized radiation, is introduced in
Sect. \ref{sect:GrainAlignment} followed by the method of deriving
synthetic Zeeman observations in Sect. \ref{sect:Zeeman}. We discuss
the resulting dust polarization pattern and LOS magnetic field
profiles in Sect. \ref{sect:ResultsDust} and
Sect. \ref{sect:ResultsZeeman}, respectively. Finally, we summarize
our results in\S~\ref{sect:Summary}.

\begin{figure*}
\begin{center}
        \begin{minipage}[c]{1.0\linewidth}
                        \begin{center}
                                \includegraphics[width=0.49\textwidth]{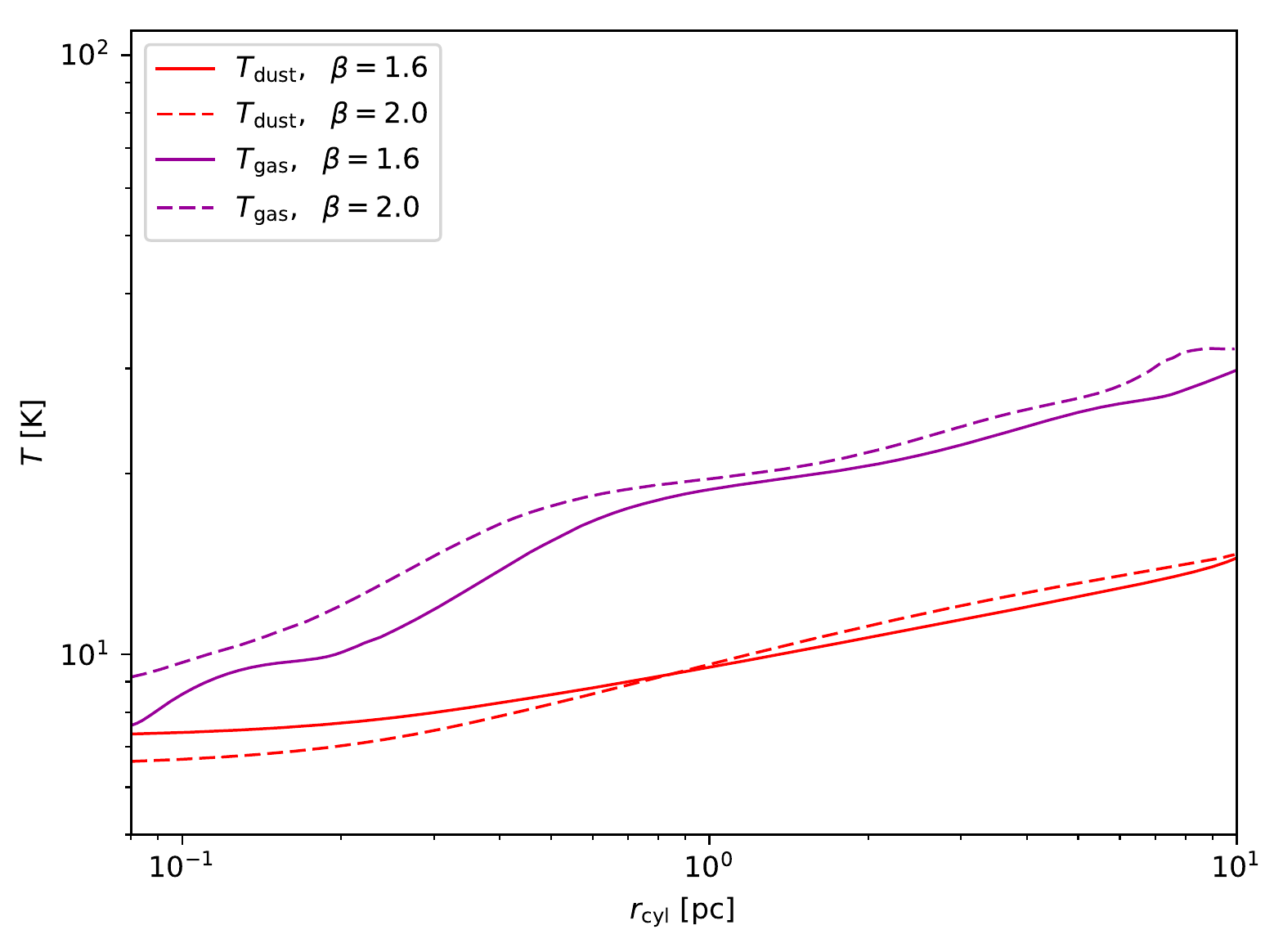}
                                \includegraphics[width=0.49\textwidth]{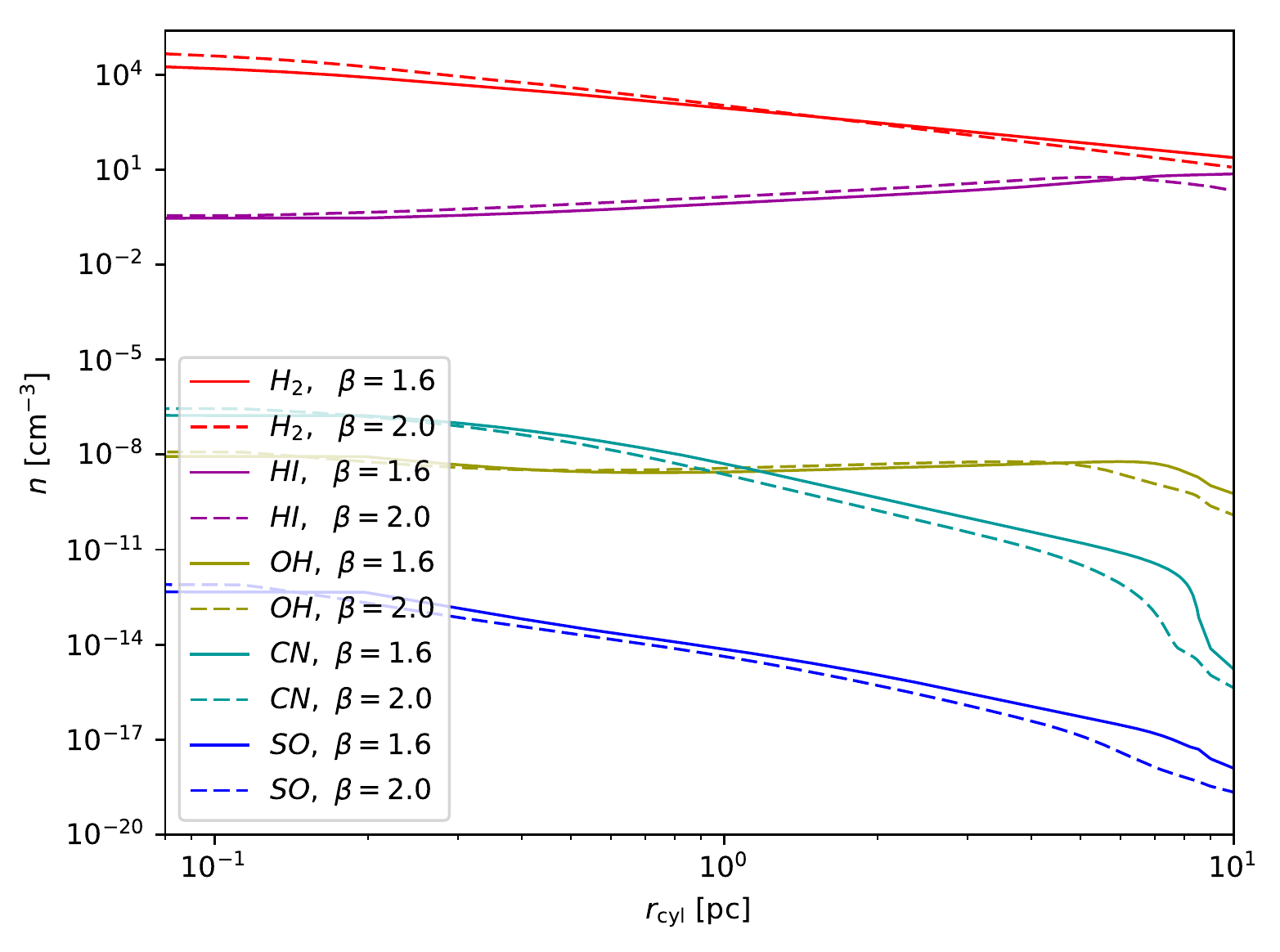}
                        \end{center}
                \end{minipage}
\end{center}
        
\caption{Left panel: Radial gas temperature $T_{\rm gas}$ distribution
  (purple) and dust temperature $T_{\rm dust}$ distribution (red) for
  a density slope index of $\beta=1.6$ (solid) and $\beta=2.0$
  (dashed). Right panel: Corresponding radial number densities for the
  molecular species considered here: $\rm{H}_2$ (red), $\rm{HI}$
  (purple), $\rm{OH}$ (yellow), $\rm{CN}$ (green), and $\rm{SO}$
  (blue) for $\beta=1.6$ (solid) and $\beta=2.0$ (dashed).  This
  figure illustrates that the parameters show above have only a weak
  dependence on the density profile power law index in
  Eqn.~\ref{eq:DensityDistribution}.}
\label{fig:TempAbundance}
\end{figure*}

\section{Filament modeling}

\subsection{Radial density and velocity profiles}
\label{sect:RadialProfiles}

Our idealized filament is modeled as a cylinder of infinite length
with its symmetry axis along the y-axis of the coordinate system (see
Fig. \ref{fig:Models}). Calculations are carried out within a cube
with a side length of $10\ \rm{pc}$ for all models. For simplicity,
the profiles of density and velocity are parametrized by the
dimensionless cylindrical radius,
\begin{equation}
|\vec{r}_{\rm cyl}|=\frac{1}{10\ \rm{pc}}\sqrt{x^2+z^2}\, ,
\end{equation}
normalized to the side length of the model.

We assume a radially symmetrical volume density distribution
consistent with \citet[][]{Stutz2016} volume density profile derived
from column density observations. We adopt the Plummer power-law
profile as suggested in \cite{Plummer1911}:
\begin{equation}
n_{\rm gas}(r_{\rm cyl}) = n_{\rm 0} \left[ 1 + \left( \frac{r_{\rm cyl}}{r_{\rm flat}}  \right)  \right]^{-\beta}\,.
\label{eq:DensityDistribution}
\end{equation}
Here, $r_{\rm flat}$ defines the characteristic radius of the density
profile close to the center of the filament where the profile becomes
flat and the parameter $\beta$ controls the slope of the density at
the outer regions. We apply a typical value in the order of $r_{\rm
  flat}= 0.05$, consistent with %\frac{1}{1\ \rm{pc}}
\citet{Palmeirim2013}, but see also \citet[][]{Smith2014,Smith2016,Boekholt2017} for the
inference of a much smaller filament volume density profile inner
softening scale. Although this Plummer profile differs from the
density profile presented in \cite{Arzoumanian2011} close to the
center we use their average value of $\beta=1.6$ as a tarting point,
which is consistent with observations in the Intergral Shaped Filement
(ISF) in Orion \citep[][]{Stutz2016}, as mentioned above and 
so we restrict our analysis here to this alignment mechanism. In order to
investigate the possible influence of different density profiles we
consider a range values of ${\beta\in[1.6:2]}$. The central number
density $n_{\rm 0}$ is chosen to keep the total mass $M_{\rm tot}$ of
the filament within the cube at a typical value of $M_{\rm tot} =
31000\ M_{\odot}$ for all parameters of $\beta$, comparable to the
mass observed in the high line mass ISF \citep{Stutz2016}.

We take that the filament is slowly contracting toward its axis of
symmetry (see Figure~\ref{fig:Models} with a velocity field defined
by:
\begin{equation}
 \vec{v}_{rad}(\vec{r}_{\rm cyl}) = -\frac{5000}{\sqrt{x^2+z^2}}\left(x,0,z\right)\ [\rm{m/s}]\,.
\end{equation}
We apply the same velocity field for all of our models if not
explicitly stated otherwise. For the line RT calculations performed in
the following sections we assume additionally a turbulent velocity
component of $v_{turb}=200\ \rm{m/s}$. The magnitude of both
velocities is chosen to be in agreement with observations
\citep[e.g.][]{Garcia2007,Arthur2016}.

\begin{figure*}
\begin{center}
 \hspace{-5.0 cm}
  \begin{minipage}[c]{0.49\linewidth}
    \begin{center}

      \includegraphics[width=1.45\textwidth]{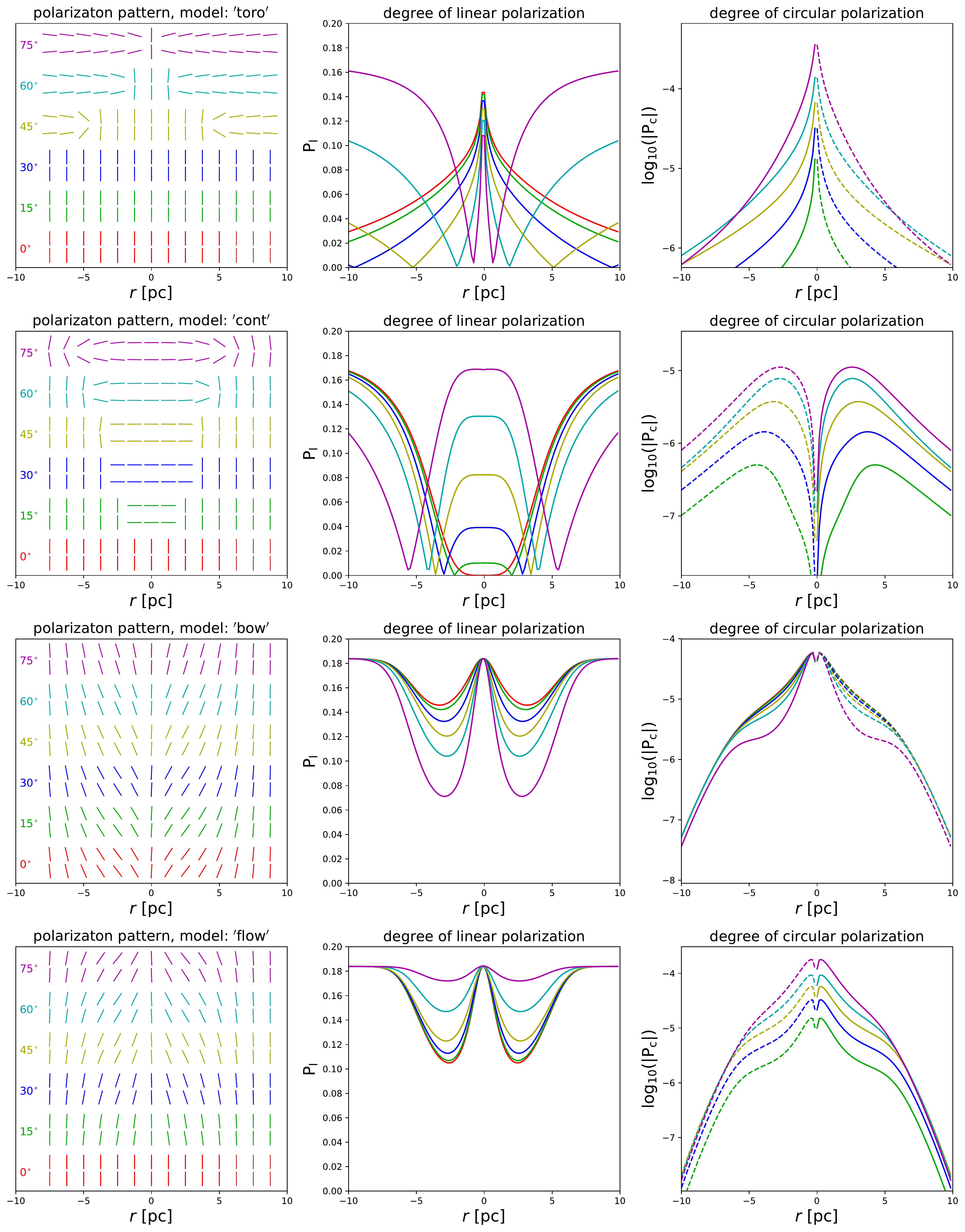}

    \end{center}
  \end{minipage}
\end{center}
\vspace{-2.0 mm}
\caption{Synthetic dust polarization quantities for the considered magnetic
  field morphologies observed at a wavelength of $\lambda=500\ \rm{\mu
    m}$. Different colors indicate results for different filament
  inclination angles ${i \in[0^{\circ},75^{\circ}]}$ in steps of
  $15^{\circ}$ as indicated by different colors, where $i = 0^{\circ}$ corresponds to viewing angle
	perpendicular to the filament axis. Left column: normalized orientation of linear
  polarization pseudo-vectors along a cut perpendicular to the
  filament axis (in the ``observer plane'', see black dashed lines in
  panel a. of Fig. \ref{fig:Models}). Middle column: degree of linear
  polarization as a function of projected radius.  Right column:
  degree of circular polarization; dashed lines represent the
  negative values of circular polarization, i.e., a flip in the
  inferred LOS field direction.}
\label{fig:PolarizationAll}
\end{figure*}

%%\newpage

\subsection{Dust grain properties}
\label{sect:GrainProperties}

\begin{figure*}
\begin{center}
 \hspace{-5.0 cm}
  \begin{minipage}[c]{0.49\linewidth}
    \begin{center}

      \includegraphics[width=1.45\textwidth]{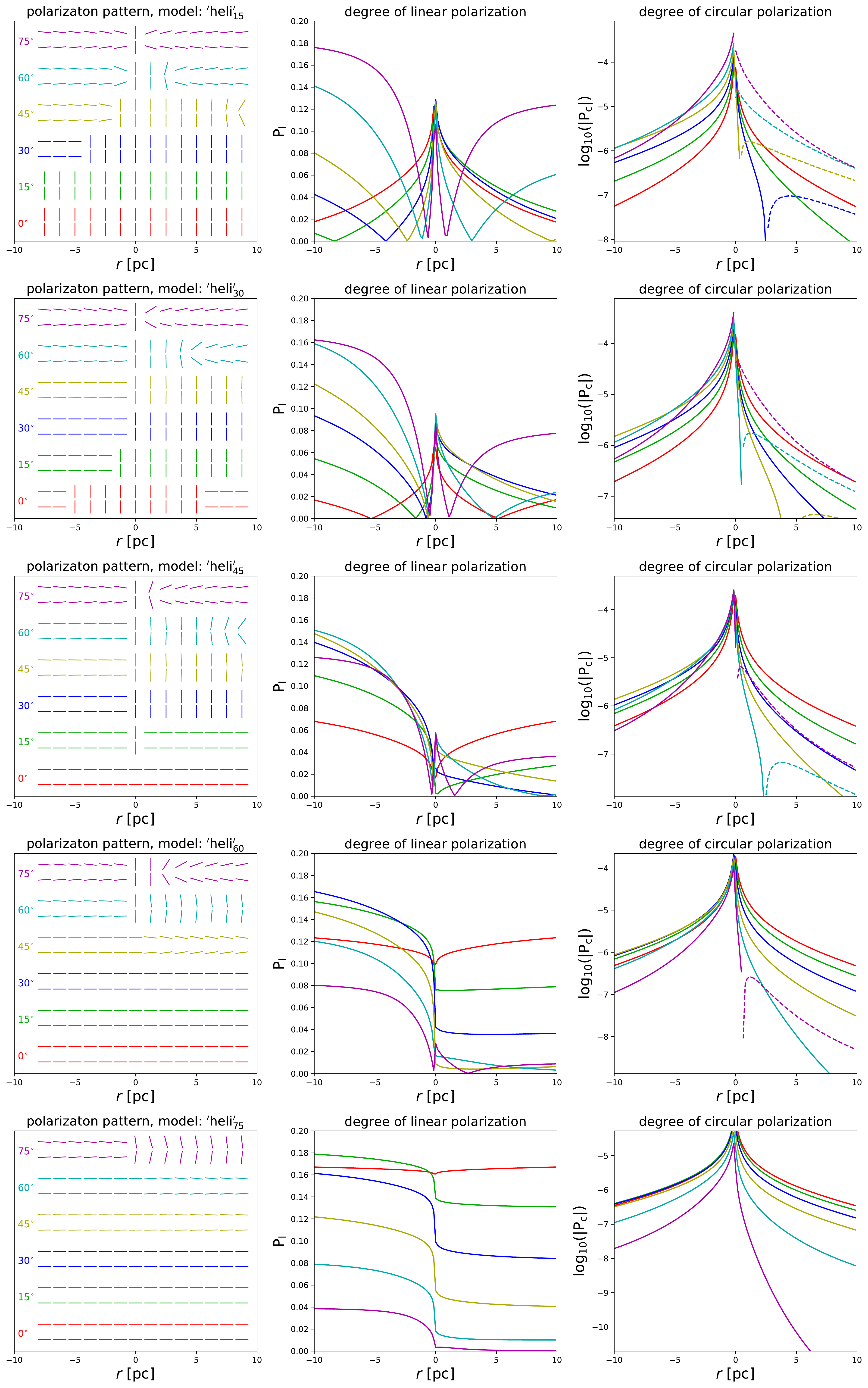}

    \end{center}
  \end{minipage}
\end{center}
\vspace{-2.0 mm}
\caption{Same as Fig.~\ref{fig:PolarizationAll} for the helical
  B-field models ($'\rm{helie}_{\rm \alpha}'$), where as the pitch angle runs  ${\alpha \in[0^{\circ},75^{\circ}]}$  in steps of $15^{\circ}$ (from top to bottom rows). As in Fig.~\ref{fig:PolarizationAll},
  different colors indicate different filament inclinations relative
  to the LOS.}
\label{fig:PolarizationHeli}
\end{figure*}

We assume dust grains to be oblate spheroids with an aspect ratio of
$0.5$ \citep[][]{Hildebrand1995, Draine2017}. The grain size is
characterized by an effective radius $a_{\rm eff }$ of a sphere of
equivalent volume. As presented in \cite{Mathis1977}, see also \cite{Weingartner2000}
for the size distribution we apply
\begin{equation}
n(a_{\rm eff }) \propto a_{\rm eff }^{-3.5}
\end{equation}
and consider a mixture of materials of $37.5\ \%$ graphite and
$62.5\ \%$ amorphous silicate grains that is consistent with the best fit
model of the extinction curve of our own galaxy \citep[][]{Mathis1977}. Although larger grain
sizes may grow in filaments, we fix the size distribution with sharp
upper and lower cut-offs at the $a_{\rm low} = 5\ \rm{nm}$ and
$a_{\rm up} = 250\ \rm{nm}$, respectively, typical for the ISM. We
apply the usual ratio of ${m_{\rm dust}/m_{\rm gas}=0.01}$ for the
dust mass to gas mas ratio \citep[][]{Mathis1977,Boulanger2000}. A
dust model with larger grains would lead to an increase in intensity
but a decrease in polarization \cite[][]{Reissl2017}. The cross
sections for extinction $C_{\rm ext, \lambda}$ and emission $C_{\rm
  abs, \lambda}$ are pre-calculated values utilizing the scattering
code DDSCAT\footnote{http://ddscat.wikidot.com/} $\rm{v7.3.2}$ \citep[][]{Draine2013} for $100$ size bins and $104$ wavelength bins
in the range of ${\lambda\in[0.9\ \rm{\mu m}: 3\ \rm{mm}]}$ \citep[see][for details]{Reissl2017}.
As input of the code we consider the
optical properties of the differently materials presented by \cite{Lee1985}
and \cite{Laor1993}. In this paper we take use of the approximation of
the efficiency factor,
\begin{equation}
      Q_{\rm{\Gamma}} = \begin{cases} 0.4   &\mbox{if } \frac{\lambda}{a_{\rm{eff}}} \leq 2 \\ 0.4\left(\frac{\lambda}{a_{\rm{eff}}} \right)^{-3}  & \mbox{if }\frac{\lambda}{a_{\rm{eff}}} > 2  \end{cases} 
        \label{eq:QGamma}
\end{equation}
that quantifies the efficiency of the dust grains to spin up when
exposed to an external an-isotropic radiation field
\citep[see][]{Lazarian2007,Hoang2014}.

\begin{figure*}
\begin{center}

        \begin{minipage}[c]{1.0\linewidth}
                        \begin{center}
                                
                                \includegraphics[width=0.489\textwidth]{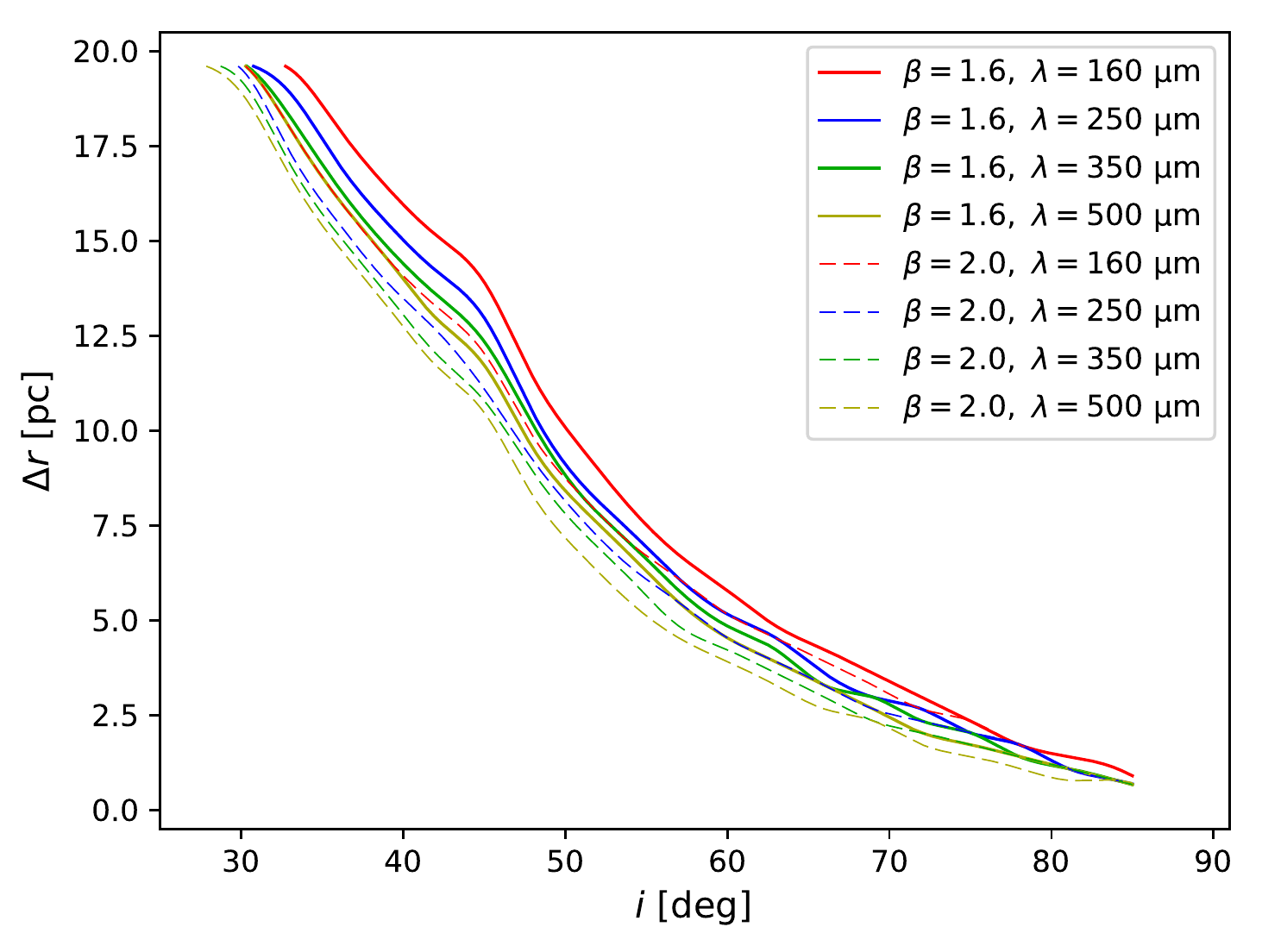}
                                \includegraphics[width=0.489\textwidth]{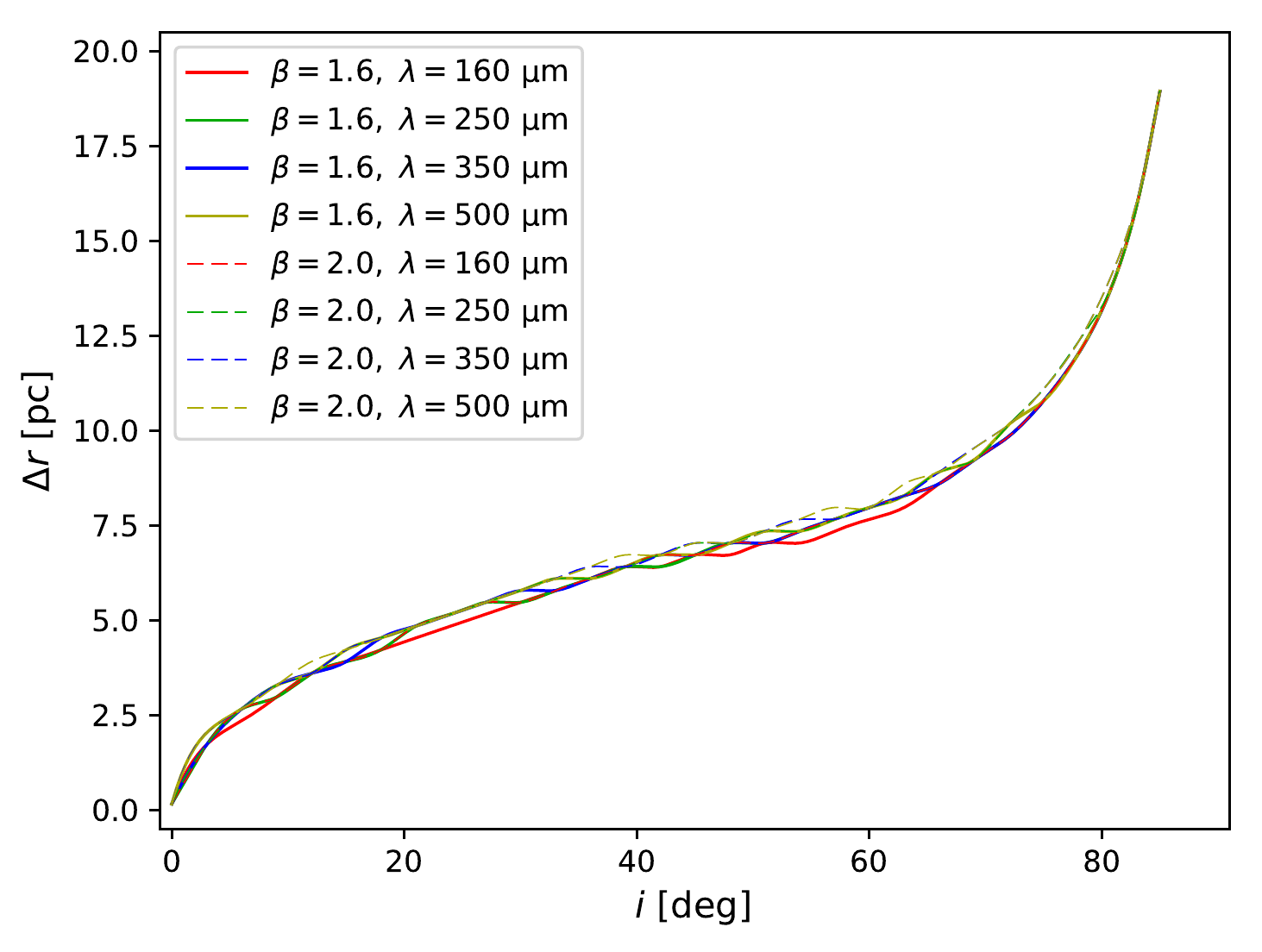}
                        \end{center}
                \end{minipage}
\end{center}
        
\caption{Left panel: Radial distances of the flipping point of the
  orientation of linear polarization as a function of inclination $i$
  for the toroidal field model ($'\rm{toro}')$ for different observed
  wavelengths $\lambda$ as well as different assumed density profile
  power law indices ($\beta$). Right panel: The same as the right
  panel for the model $'\rm{cont}'$.}
\label{fig:DistanceToroCont}

\begin{center}

        \begin{minipage}[c]{1.0\linewidth}
                        \begin{center}
                                \includegraphics[width=0.489\textwidth]{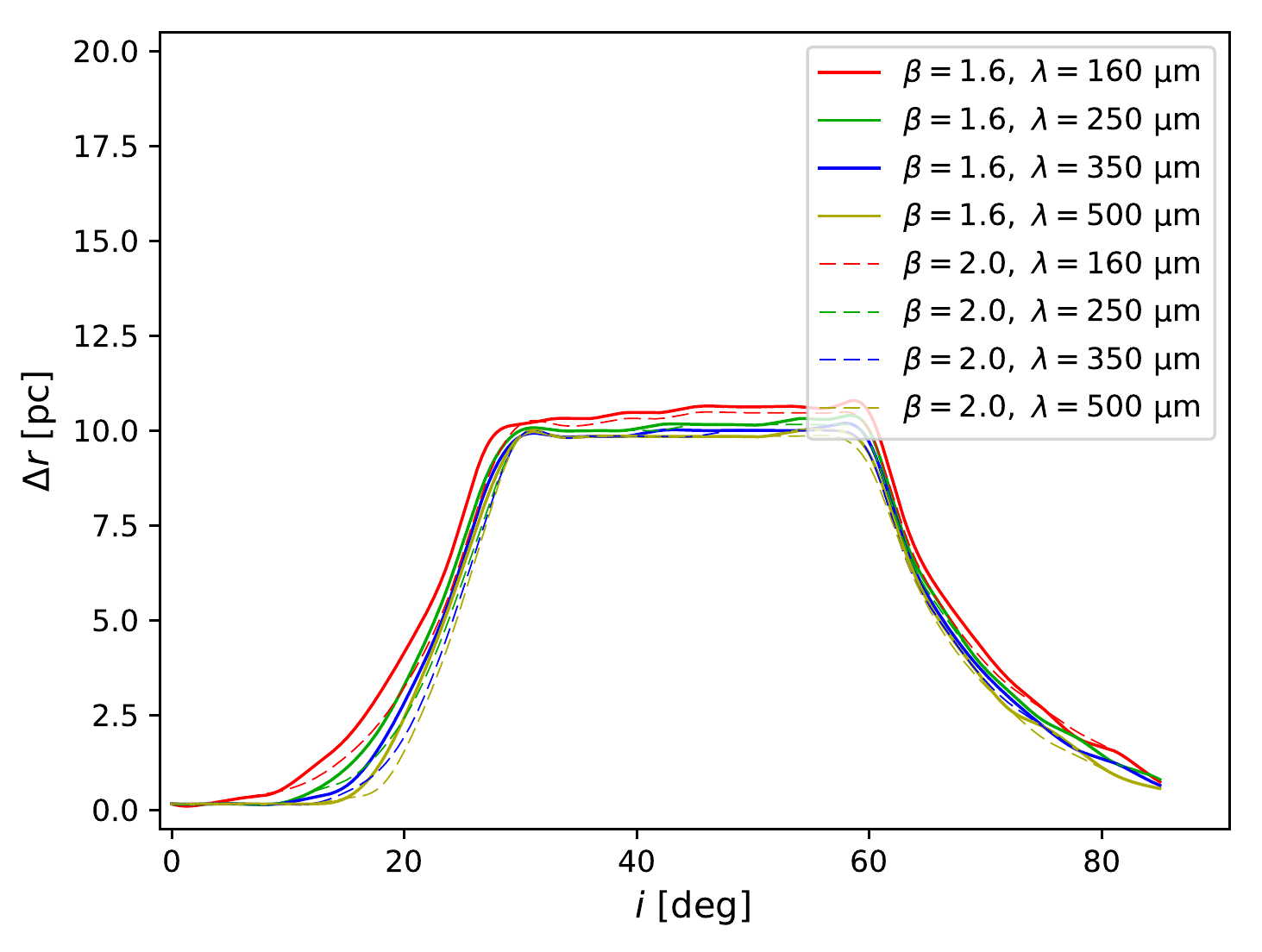}
                                \includegraphics[width=0.489\textwidth]{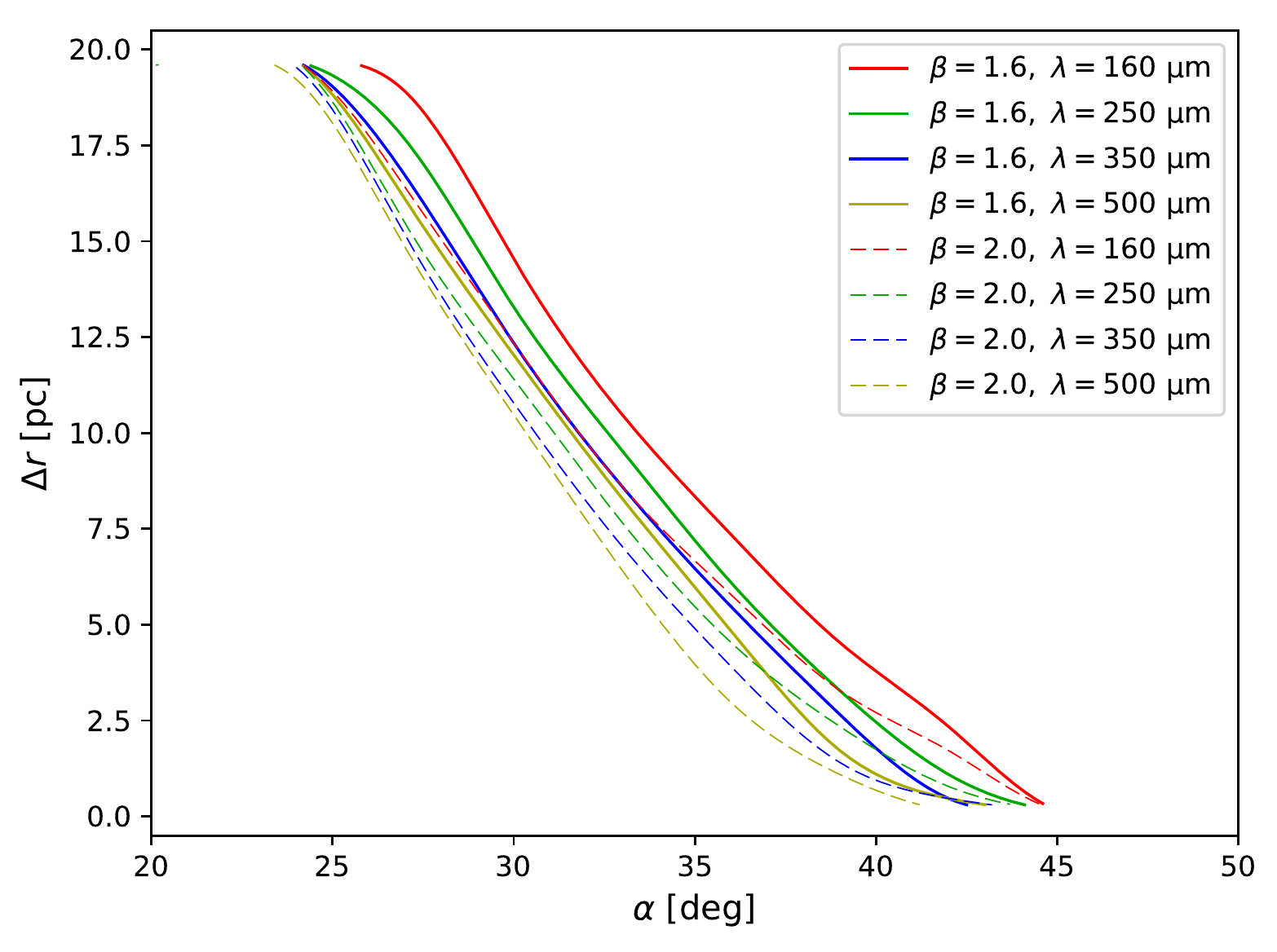}
                        \end{center}
                \end{minipage}
\end{center}
        
\caption{Left panel: The same as Fig. \ref{fig:DistanceToroCont} for the
  model $'\rm{heli}_{\rm 45}'$ with a fixed pitch angle of $\alpha=45^{\circ}$ and an inclination $i\in[3^{\circ},87^{\circ}]$. Right panel: Radial distance of the
  flipping points for the class of helical models $'\rm{heli}_{\rm
    \alpha}'$ as a function of pitch angle $\alpha$ for a fixed  inclination of $i=0^{\circ}$.}
\label{fig:DistanceHeli}
\end{figure*}

\begin{figure*}
\begin{center}
        \begin{minipage}[c]{0.9\linewidth}
                        \begin{center}
                                \includegraphics[width=0.49\textwidth]{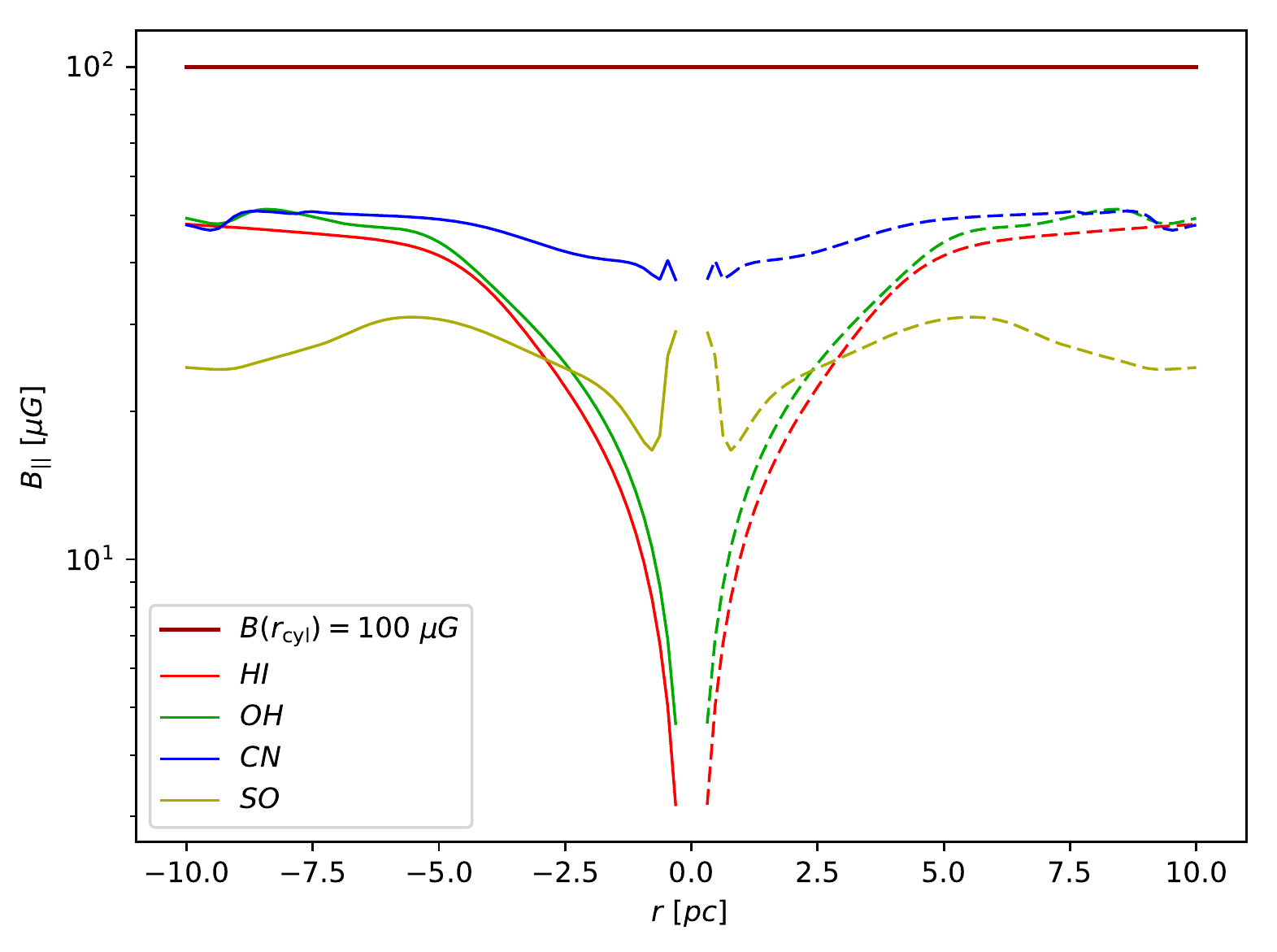}
                                \includegraphics[width=0.49\textwidth]{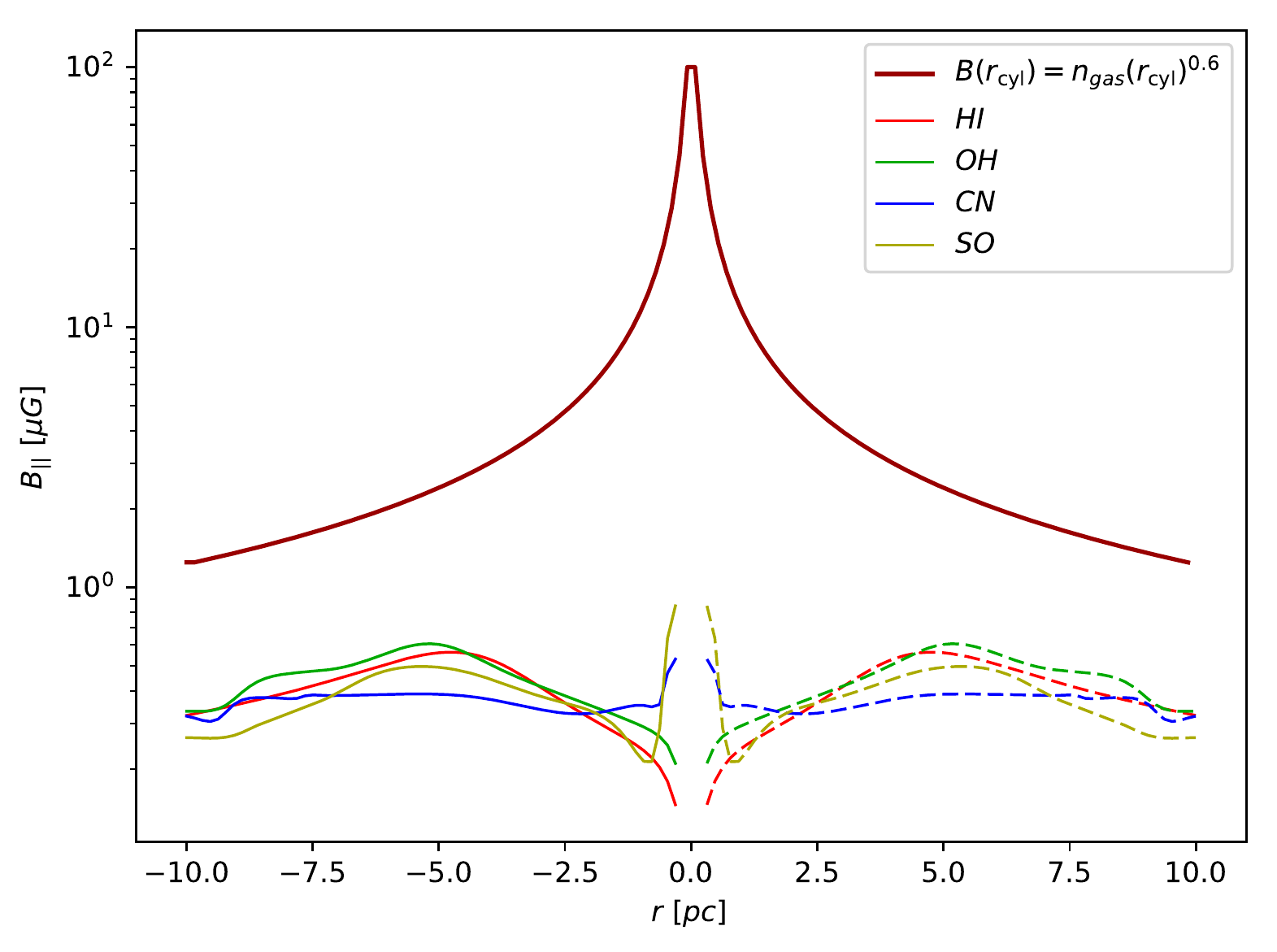}
                        \end{center}
        \end{minipage}
       
       \begin{minipage}[c]{0.9\linewidth}
                        \begin{center}
                                \includegraphics[width=0.49\textwidth]{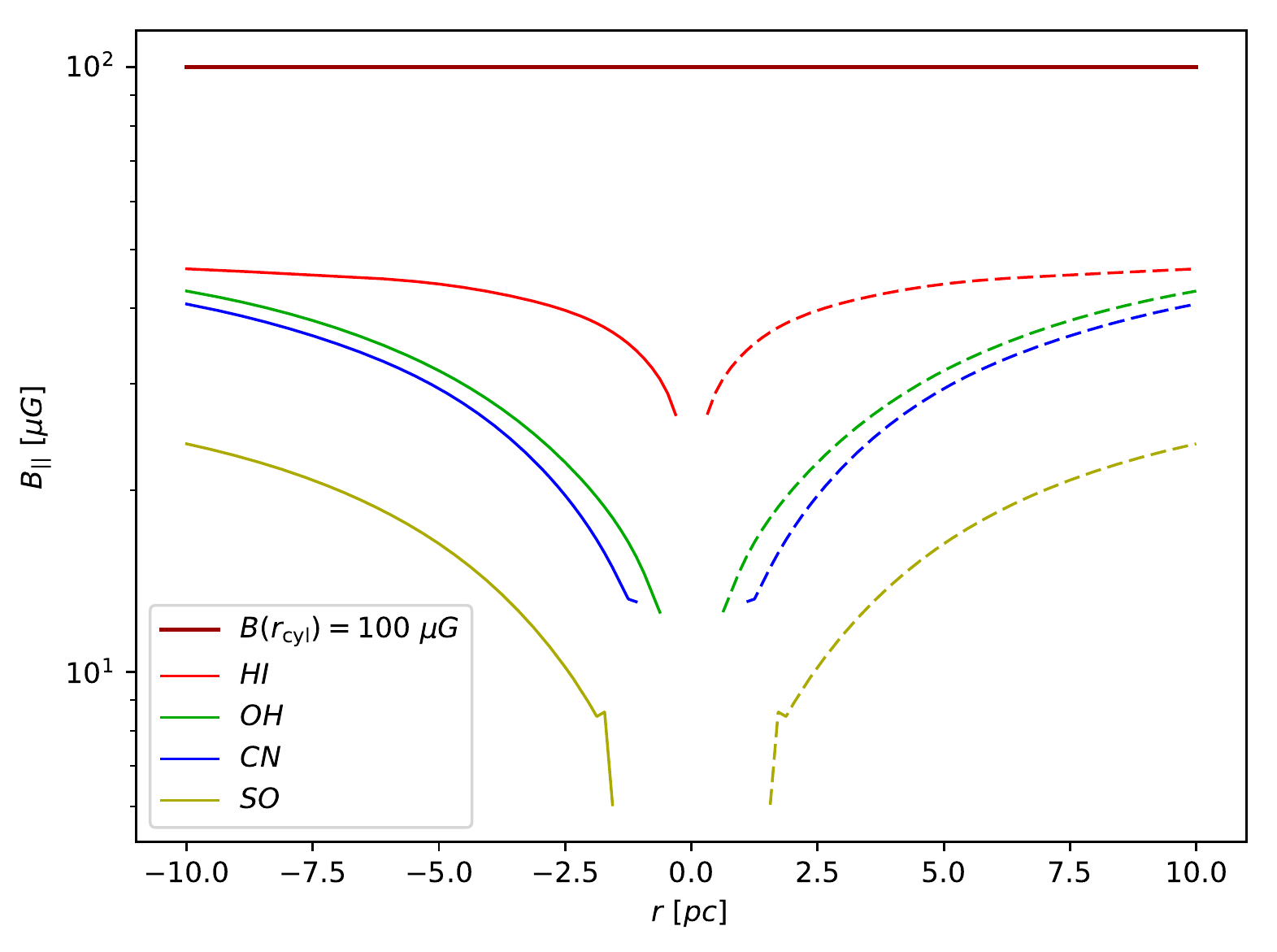}
                                \includegraphics[width=0.49\textwidth]{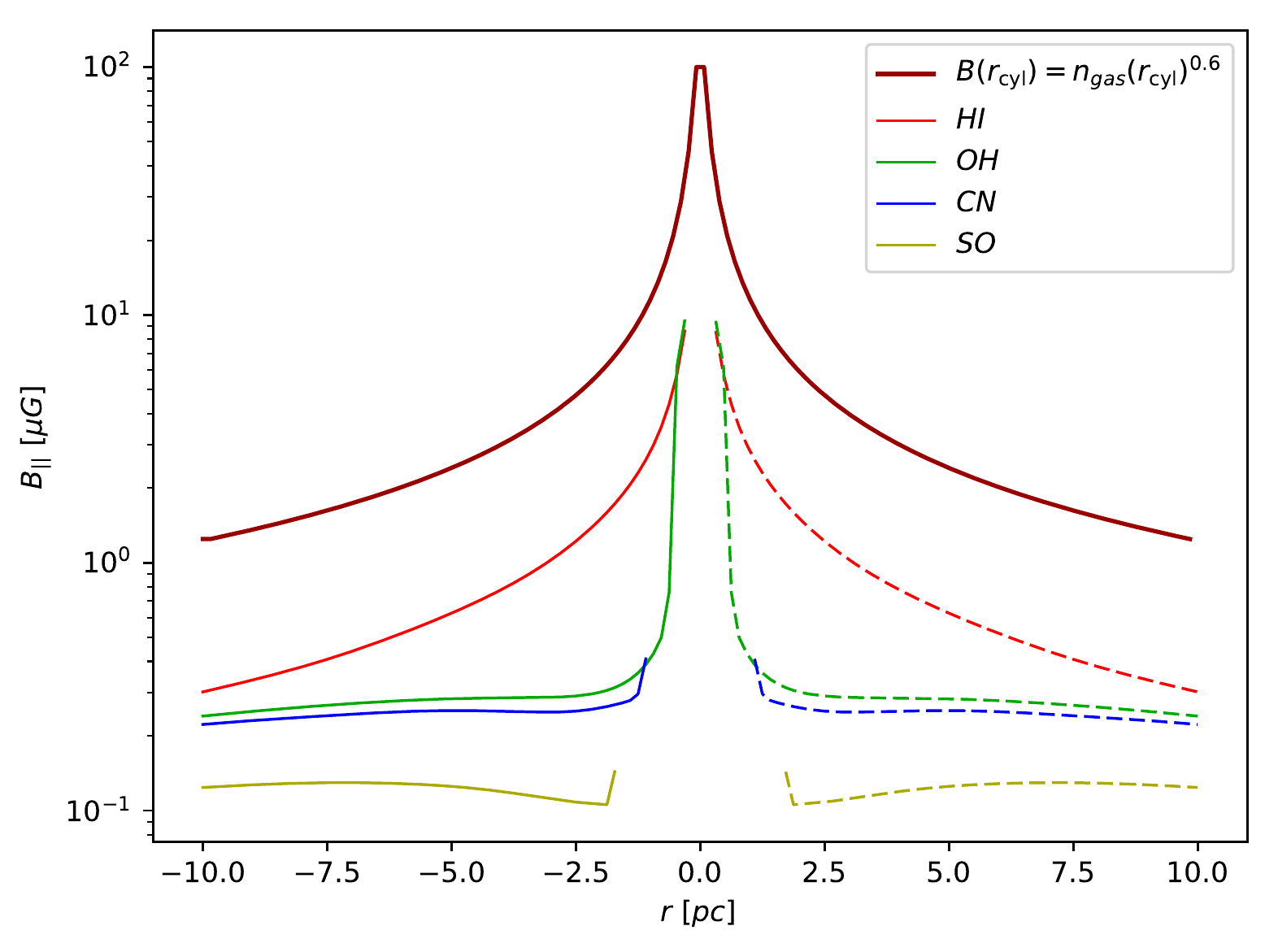}
                        \end{center}   
        \end{minipage}
\end{center}
        
\caption{Assumed radial 3D dependence of the field strength $B(r_{\rm
    cyl})$ (thick dark red line) compared to the inferred LOS magnetic
  field strength $B_{\rm ||}$ derived from synthetic Zeeman
  measurements for the toroidal model for an inclination of $i=0^{\circ}$. Different molecular tracer Zeeman
  results are shown (color coded). Left column: We assume a
  constant field strength of ${B_{\rm 0}(r_{\rm cyl})=100\ \rm{\mu
      G}}$. Right column: $B_{\rm 0}(r_{\rm cyl})$ scales with the gas
  density (see Eqn.~\ref{eq:Brad}).  Top row: derived magnetic fields
  calculated with the simulated abundances shown in
  Fig.~\ref{fig:TempAbundance}.  Bottom row: derived magnetic fields
  calculated assuming a constant ratio of $n/n_{\rm gas} = 10^{-6}$. Solid lines indicate positive values and dashed lines negative values of $B_{\rm ||}$.}
\label{fig:ZeemanToro}
\end{figure*}

%\begin{figure*}
%\begin{center}
        %\begin{minipage}[c]{0.9\linewidth}
                        %\begin{center}
                                %\includegraphics[width=0.49\textwidth]{mag_heli45_20_c.pdf}
                                %\includegraphics[width=0.49\textwidth]{mag_heli45_20_v.pdf}
                        %\end{center}
        %\end{minipage}
       %
       %\begin{minipage}[c]{0.9\linewidth}
                        %\begin{center}
                                %\includegraphics[width=0.49\textwidth]{mag_heli45_20_c_const.pdf}
                                %\includegraphics[width=0.49\textwidth]{mag_heli45_20_v_const.pdf}
                        %\end{center}   
        %\end{minipage}
%\end{center}
%\caption{The same as Fig. \ref{fig:ZeemanToro} for the model
  %$'\rm{heli}_{\rm 45}'$.}
%\label{fig:ZeemanHeli45}
%\end{figure*}

\begin{figure*}
\begin{center}
        \begin{minipage}[c]{0.9\linewidth}
                        \begin{center}
                                \includegraphics[width=0.49\textwidth]{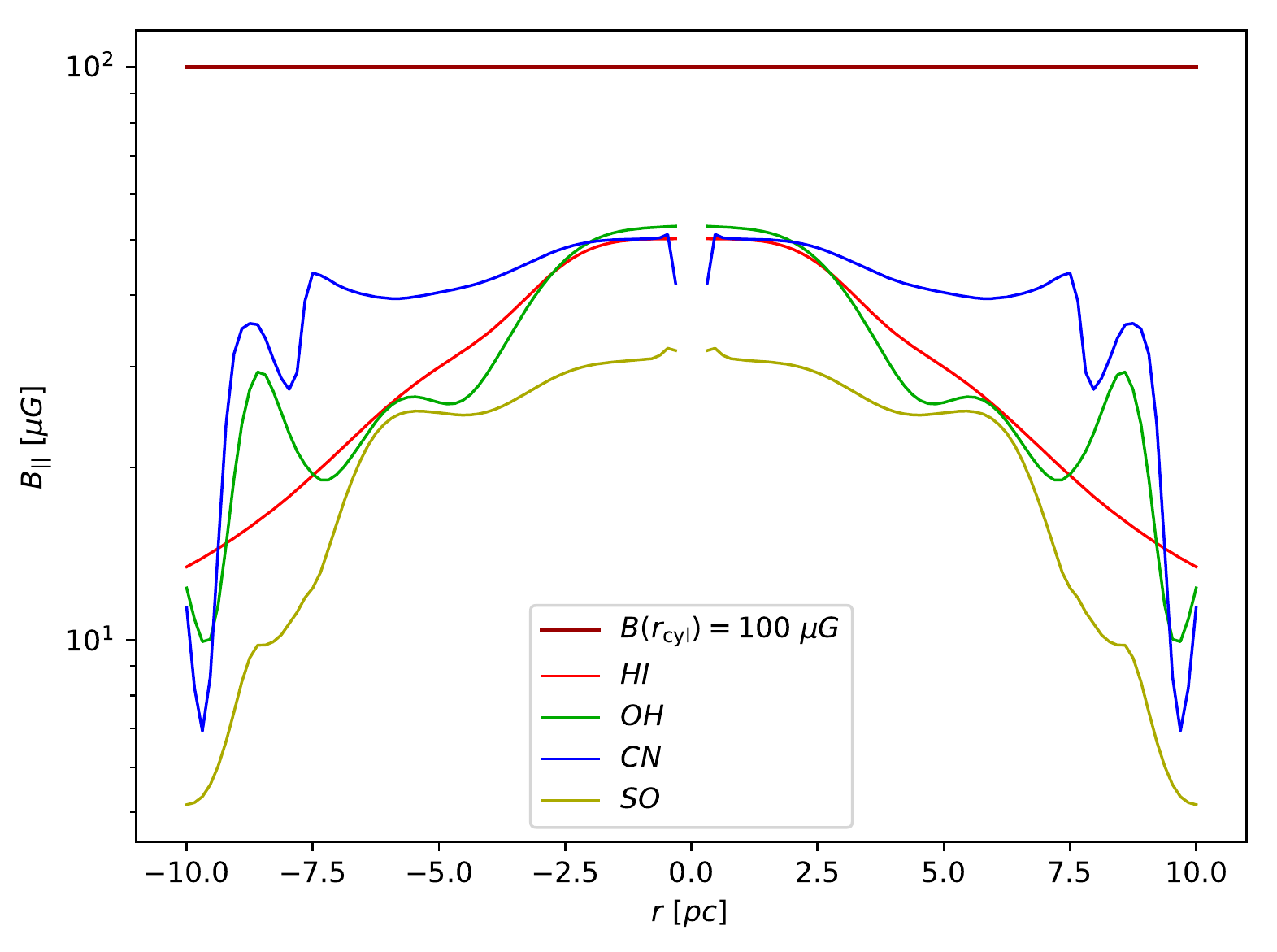}
                                \includegraphics[width=0.49\textwidth]{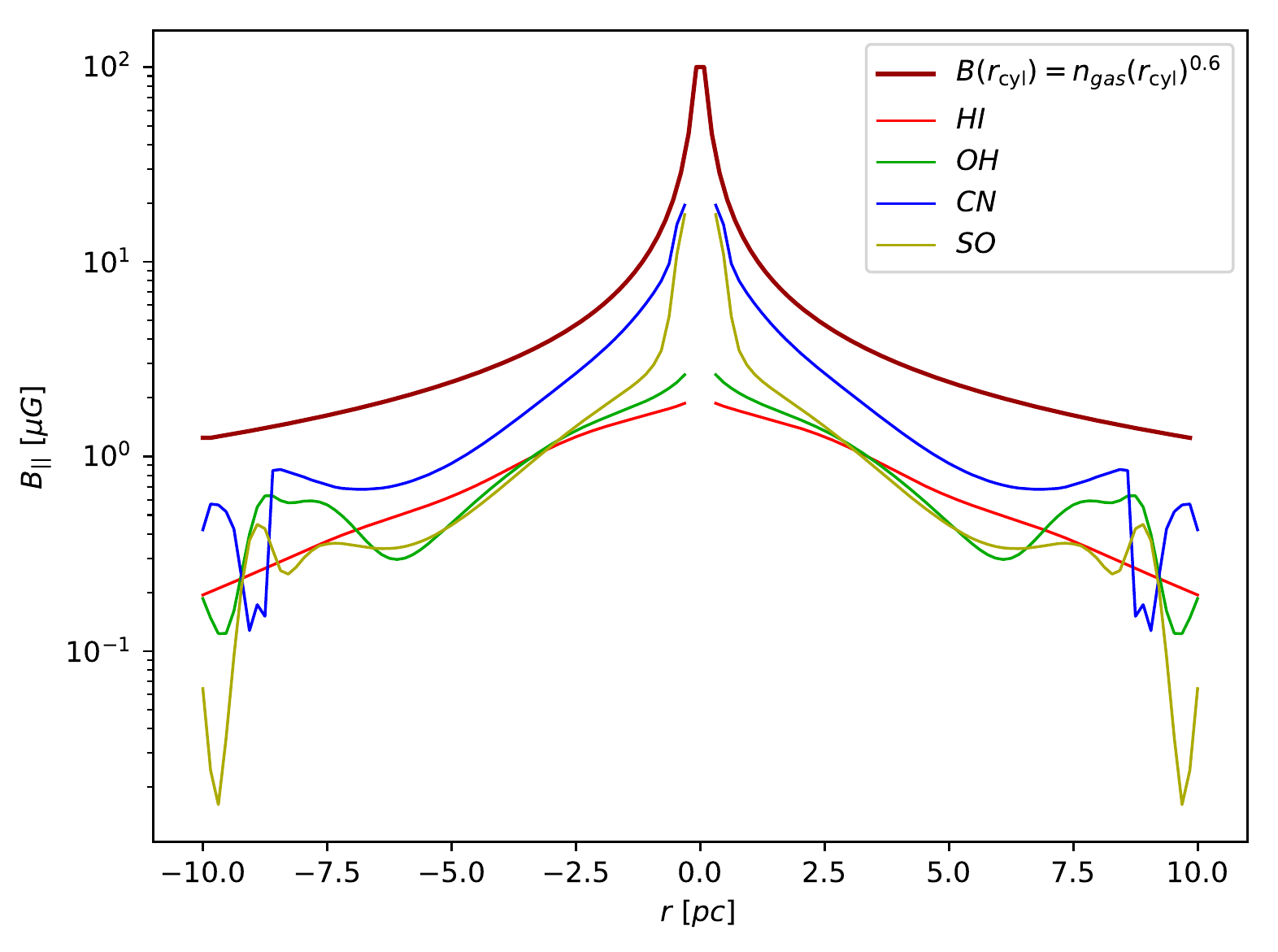}
                        \end{center}
        \end{minipage}
       
       \begin{minipage}[c]{0.9\linewidth}
                        \begin{center}
                                \includegraphics[width=0.49\textwidth]{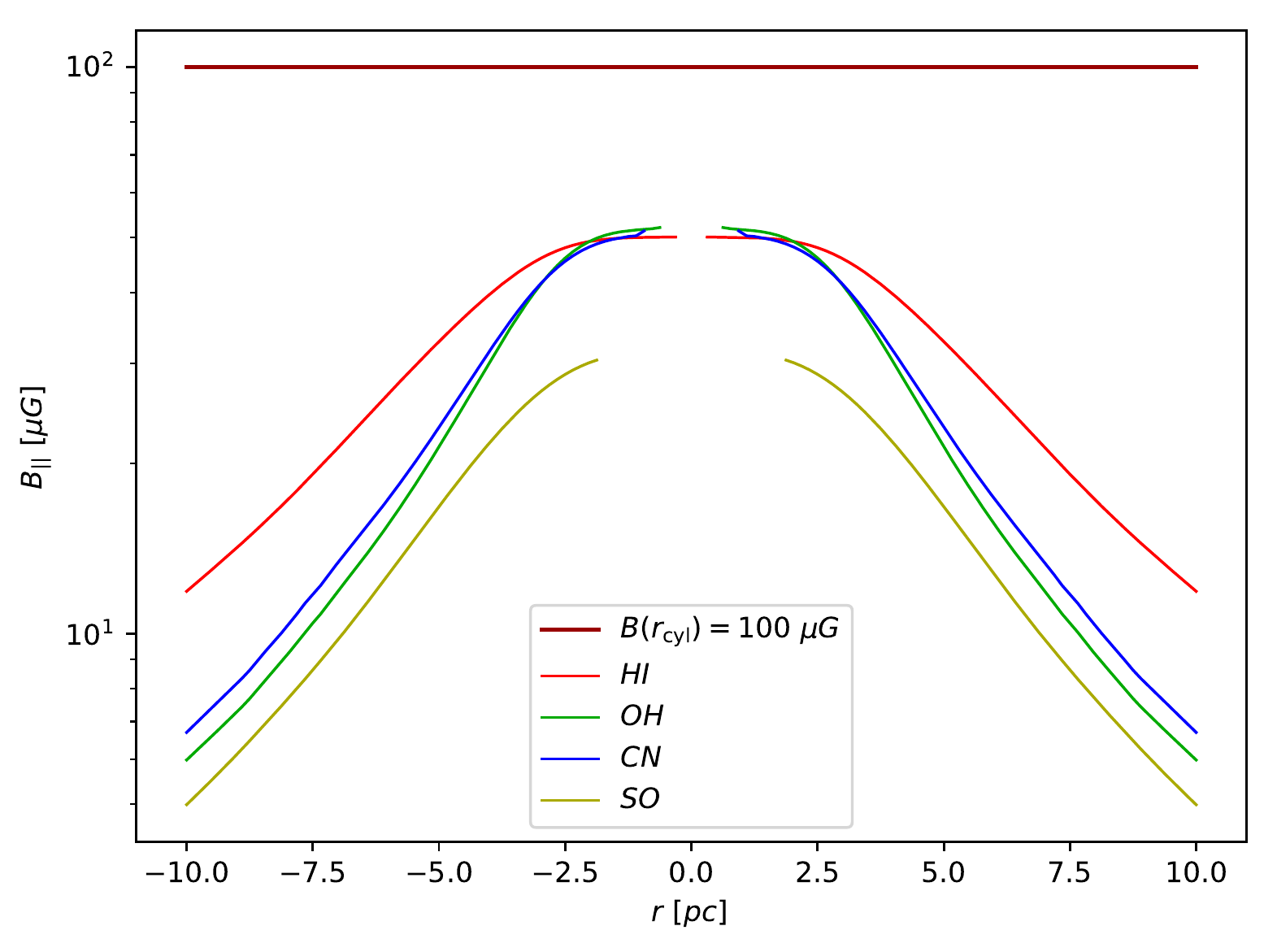}
                                \includegraphics[width=0.49\textwidth]{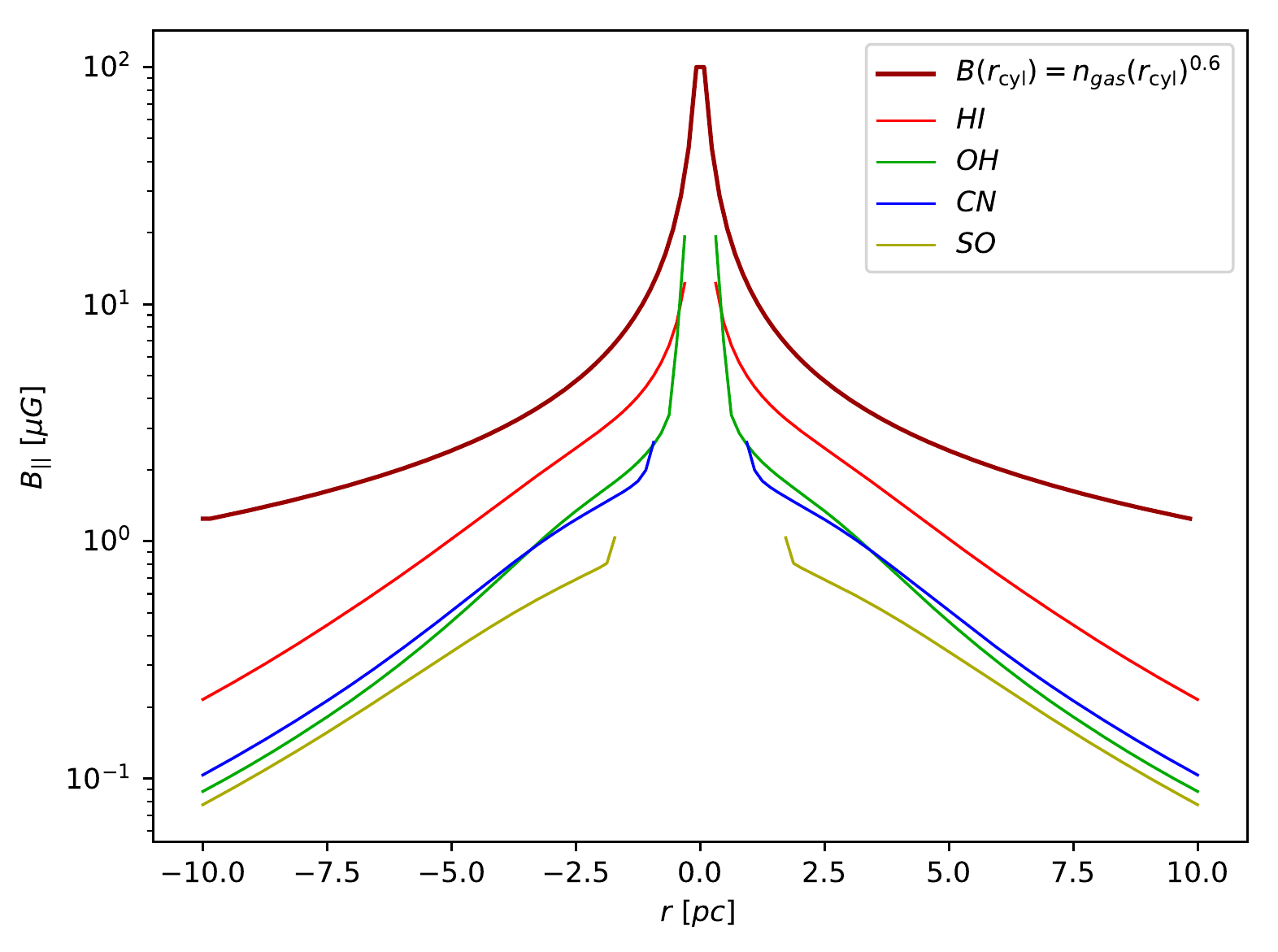}
                        \end{center}   
        \end{minipage}
\end{center}        
\caption{The same as Fig. \ref{fig:ZeemanToro} for the model
  $'\rm{cont}'$.}
\label{fig:ZeemanCont}
\end{figure*}

\begin{figure*}
\begin{center}
        \begin{minipage}[c]{0.9\linewidth}
                        \begin{center}
                                \includegraphics[width=0.49\textwidth]{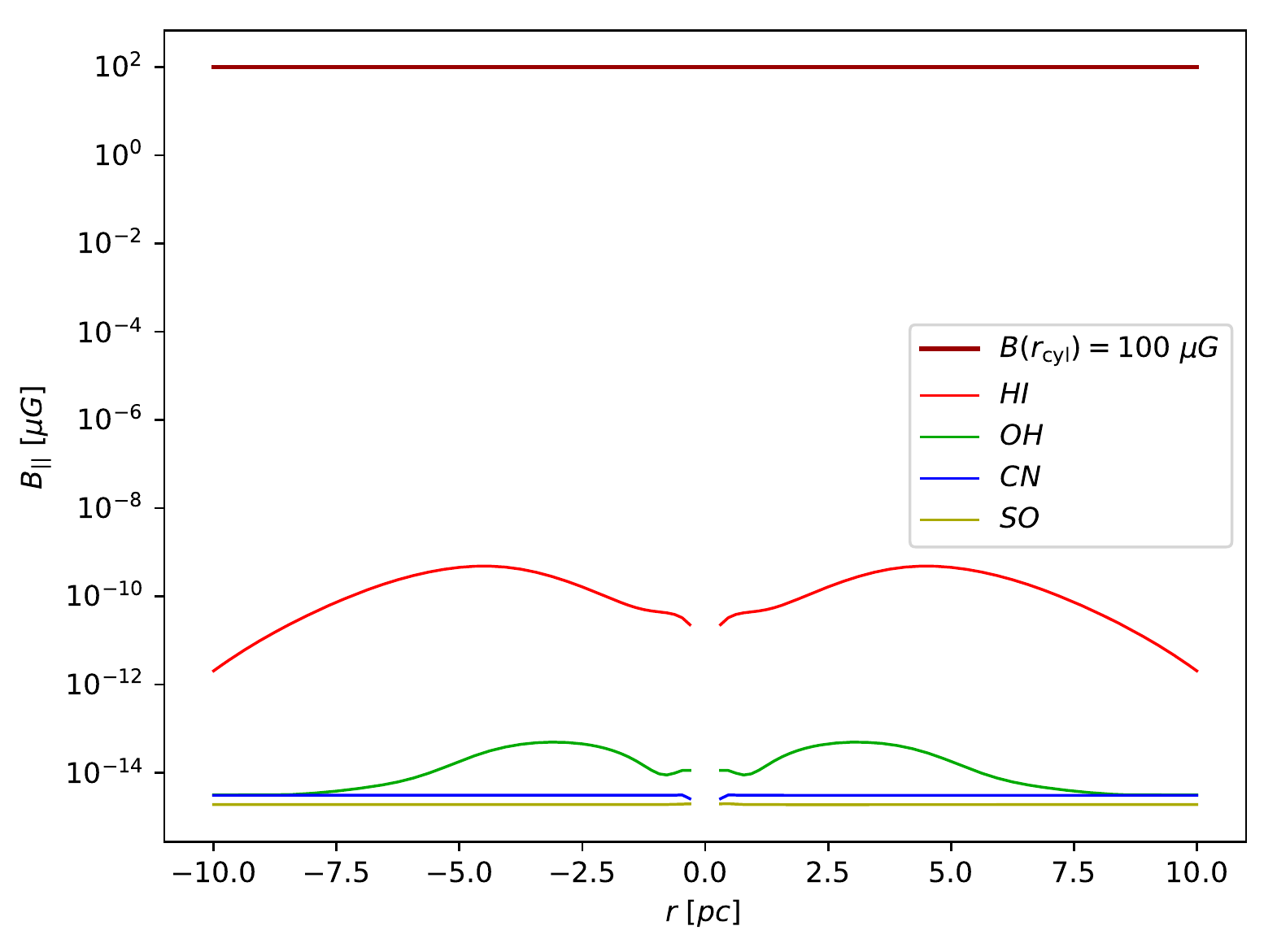}
                                \includegraphics[width=0.49\textwidth]{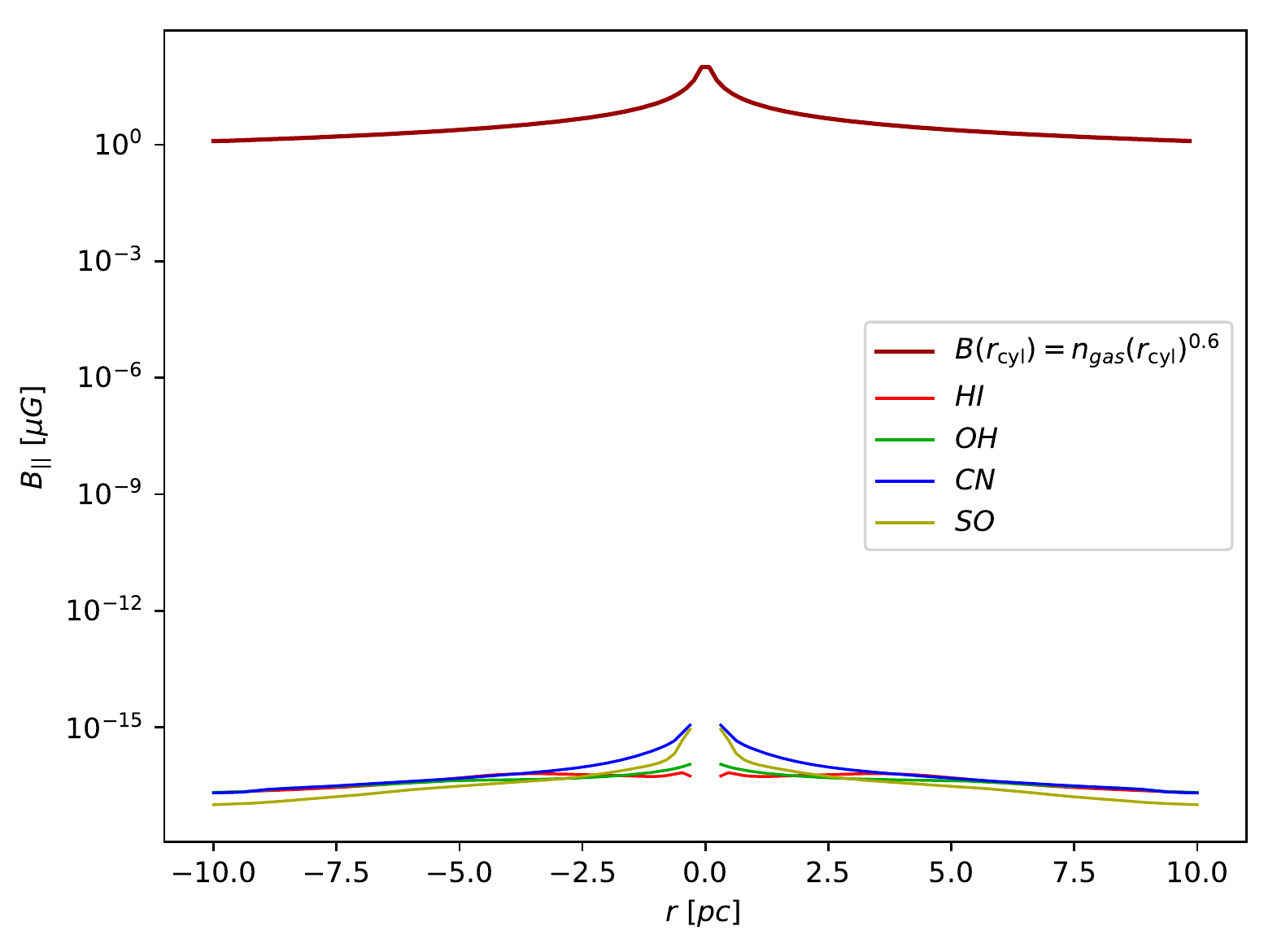}
                        \end{center}
        \end{minipage}
       
       \begin{minipage}[c]{0.9\linewidth}
                        \begin{center}
                                \includegraphics[width=0.49\textwidth]{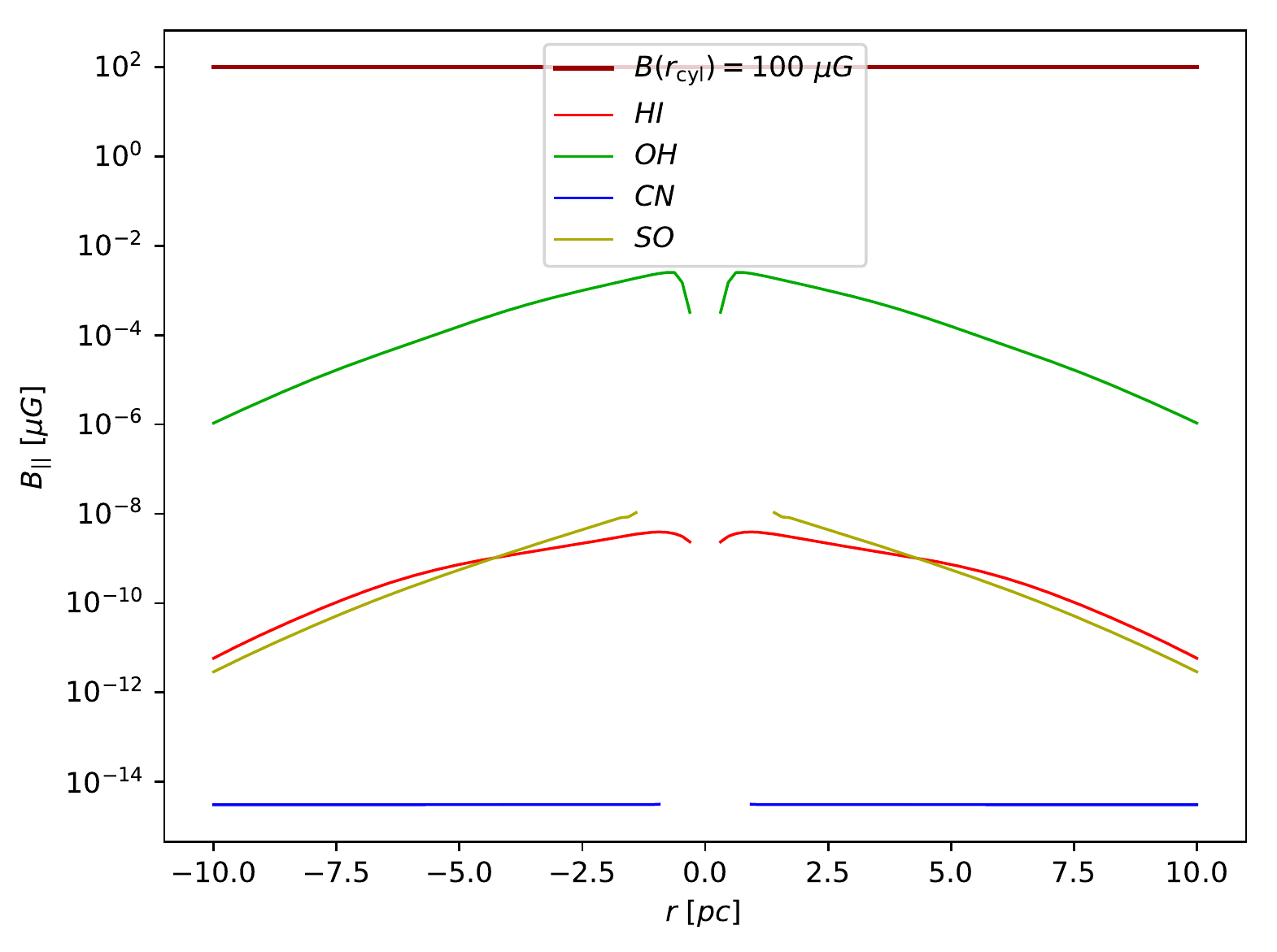}
                                \includegraphics[width=0.49\textwidth]{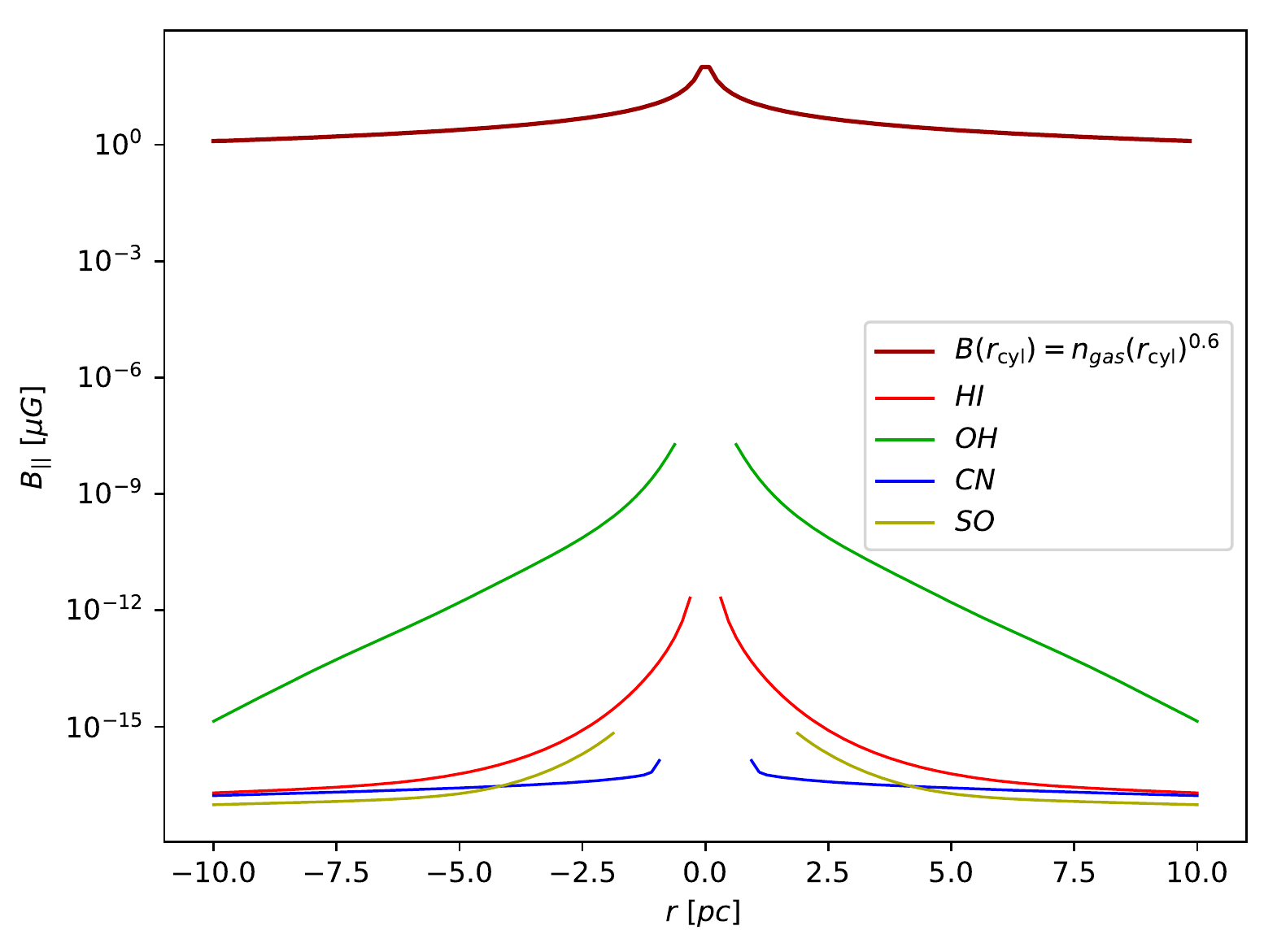}
                        \end{center}
       \end{minipage}       
\end{center}        
\caption{The same as Fig. \ref{fig:ZeemanToro} for the model
  $'\rm{bow}'$.}
\label{fig:ZeemanBow}
\end{figure*}

\begin{figure*}
\begin{center}
        \begin{minipage}[c]{0.9\linewidth}
                        \begin{center}
                                \includegraphics[width=0.49\textwidth]{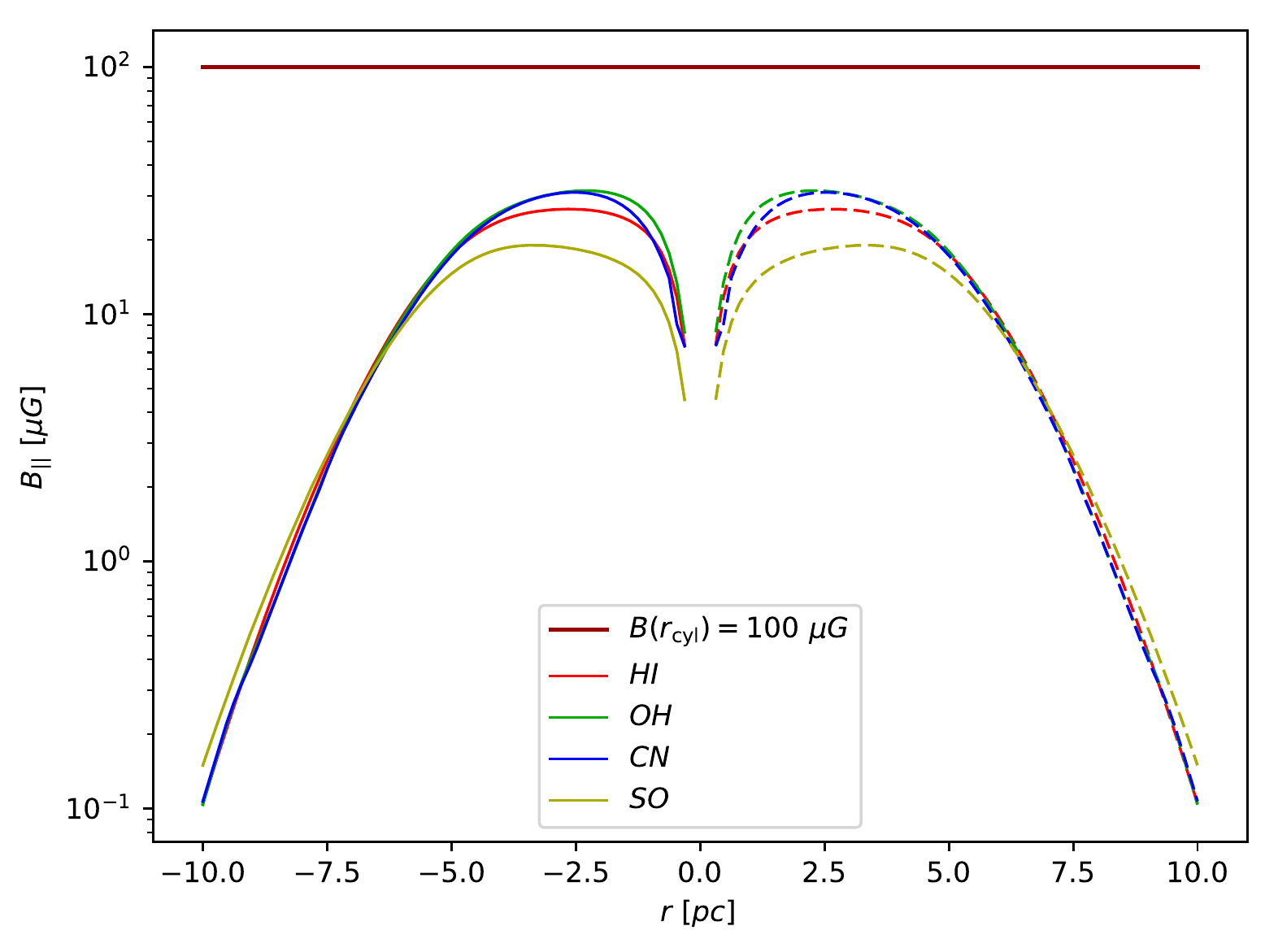}
                                \includegraphics[width=0.49\textwidth]{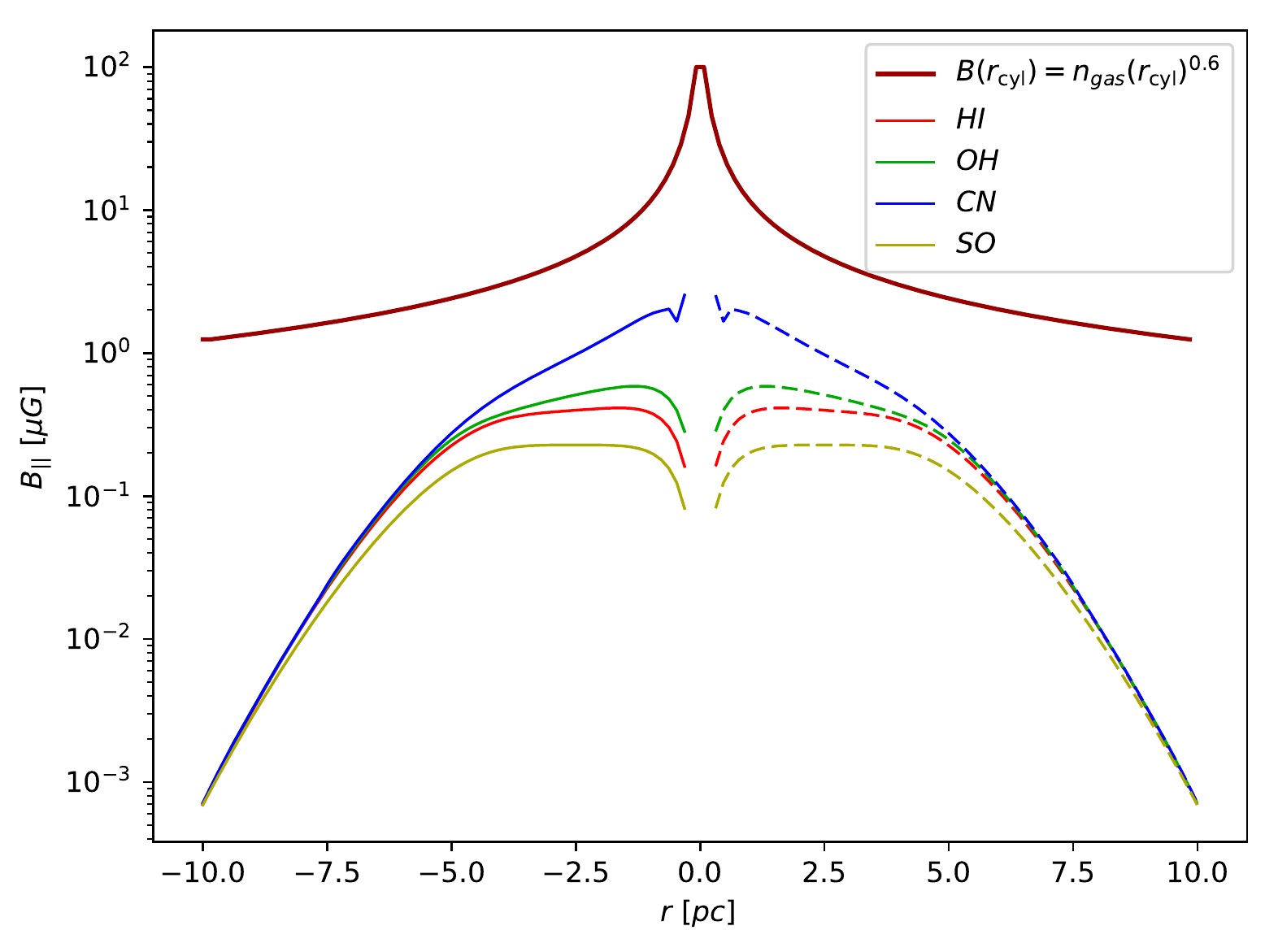}
                        \end{center}
        \end{minipage}
       
       \begin{minipage}[c]{0.9\linewidth}
                        \begin{center}
                                \includegraphics[width=0.49\textwidth]{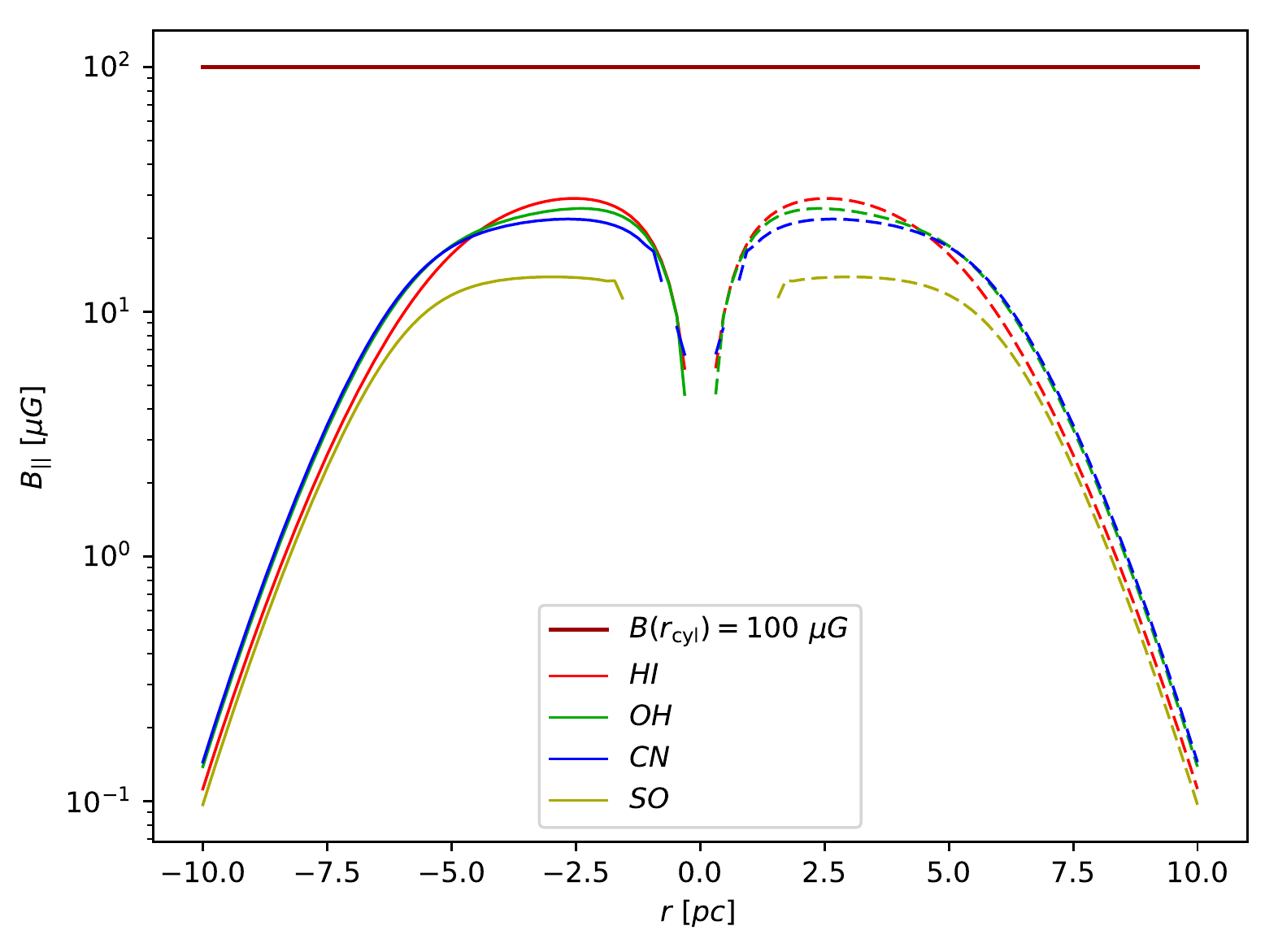}
                                \includegraphics[width=0.49\textwidth]{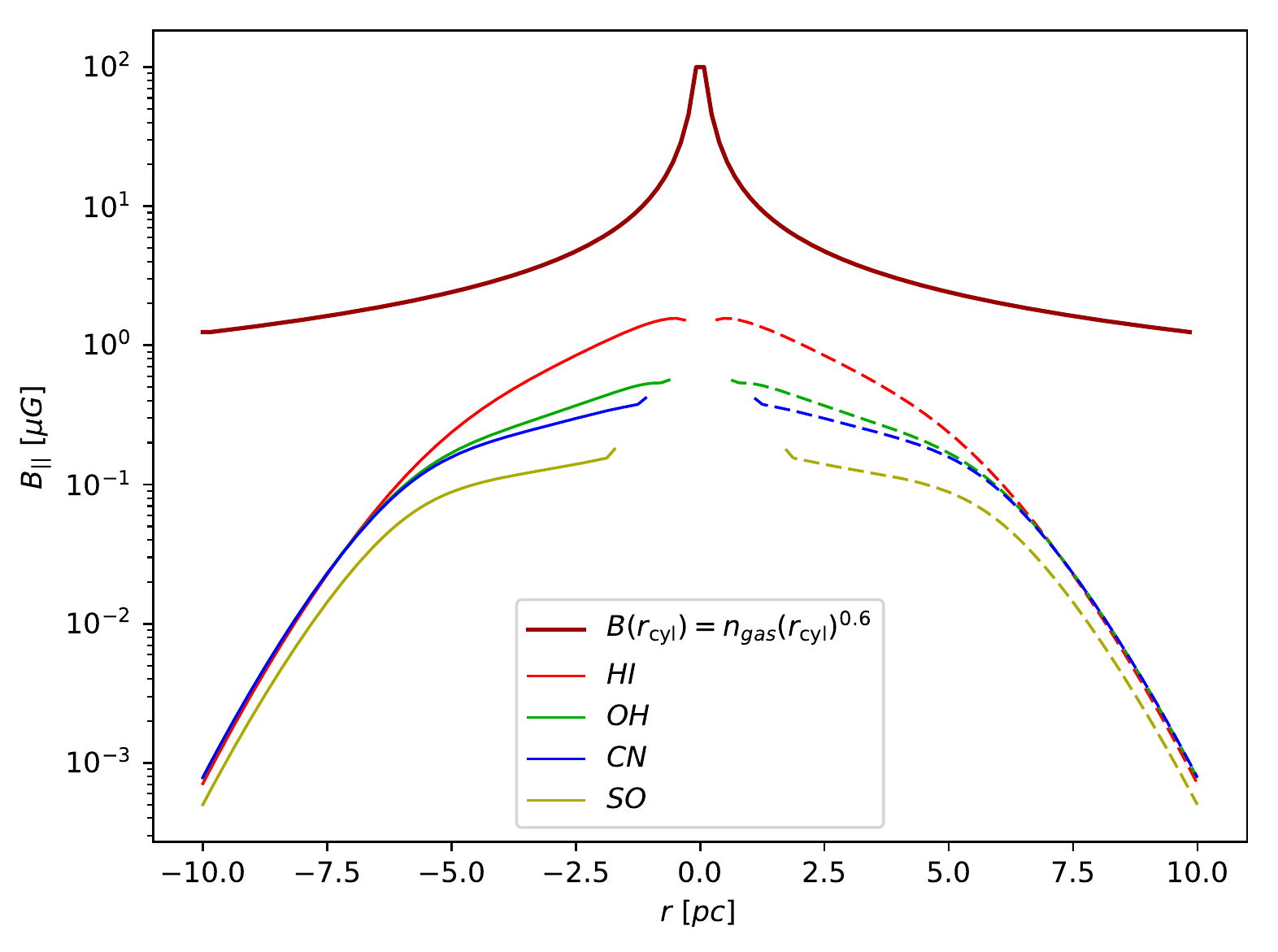}
                        \end{center}   
        \end{minipage}
\end{center}
        
\caption{The same as Fig. \ref{fig:ZeemanToro} for the model
  $'\rm{flow}'$.}
\label{fig:ZeemanFlow}
\end{figure*}

\subsection{Magnetic field morphologies}
\label{sect:FieldMorphologies}

Several models have been proposed over the years arguing for the
stability and shape of filaments on the basis of toroidal or helical
fields \citep[e.g.,][]{Nagasawa1987,Fiege2000a,Toci2015b}. However,
observational constraints by means of either dust polarization or
Zeeman measurements for such fields have been difficult to interpret
\citep[e.g.,][]{heiles1997,falgarone2001,Palmeirim2013}. Hence, we
model a purely toroidal by an analytic expression with

\begin{equation}
\vec{B}(\vec{r}_{\rm cyl}) =   \frac{B_{\rm 0}(r_{\rm cyl})}{\sqrt{x^2+z^2}}  \left(-z,0,x\right)^T\, ,
\label{eq:MagToro}
\end{equation}
where $B_{\rm 0}(r_{\rm cyl})$ is a function accounting for the radial
magnetic field strength (see below) and the superscript $T$ stands for a transposed vector. 
We label this kind of model as $'\rm{toro}'$ in the following sections.
However, a purely toroidal field is just a special case in the much
broader class of helical fields. Helical magnetic fields can
conveniently be modeled by
\begin{equation}
\begin{split}
\vec{B}(\vec{r}_{\rm cyl}) = \qquad\qquad\qquad\qquad\qquad\qquad\qquad\qquad\qquad\qquad\qquad \\  B_{\rm 0}(r_{\rm cyl}) \left(-z\cos(\alpha)/\sqrt{x^2+z^2},\sin(\alpha),x\cos(\alpha)/\sqrt{x^2+z^2}\right)^T\,.
\end{split}
\label{eq:MagHeli}
\end{equation}
Here, $\alpha$ is the pitch-angle of the field where
$\alpha=0^{\circ}$ represents the toroidal case above. We refer to
this class of models as $'\rm{heli}_{\rm \alpha}\ '$ and consider a
range of ${\alpha\in[3^{\circ}:87^{\circ}]}$. A 3D representation of the
toroidal and helical, respectively, field can be found in panels (b)
and (c) of Fig.~\ref{fig:Models}.
As the filament moves with respect to the environment, the gas mass
will affect the magnetic field morphology. The strength of this effect
heavily depends on the trajectory of the gas. Moving gas contracting
along magnetic field lines does not influence the magnetic field
morphology, while a perpendicular contraction may bend the field
symmetrically with respect to the y-axis. Assuming the field is
initially parallel to the z-axis in Fig.~\ref{fig:Models}, such a
field morphology can be modeled by the following expression:
\begin{equation}
\vec{B}(\vec{r}_{\rm cyl}) =   \frac{B_{\rm 0}(r_{\rm cyl}) \left(\rm{sgn}(x) 30zx^2\exp(-2z^2),0,1\right)^T}{1+900z^2x^4\exp(-4z^2)}\,.
\label{eq:MagCont}
\end{equation}
As the magnetic field would abruptly switch sign when x goes from negative to positive, the term $\rm{sgn}(x)$ ensures the continuity of the field. We refer to this kind of morphology emerging from a
contracting filament as model $'\rm{cont}'$; see also panel (d) in Fig.~\ref{fig:Models}.
Contraction of mass is not limited to a mass flow perpendicular to the
symmetry axis. As indicated by numerical simulations
\citep[][]{Gomez2018}. Ther could also be contraction along the filament as in the models by \cite{Smith2016}. In our model as well as in observations \citep[e.g.][]{Pattle2017} a collapse may also take place along the
filament.  In this case the magnetic field would be warped along the
predominant trajectory of the mass, as in panel (e) of
Fig.~\ref{fig:Models}. We model the gross geometric characteristics of
such a field with a radially dependent Gaussian:
\begin{equation}
\vec{B}(\vec{r}_{\rm cyl}) =   \frac{B_{\rm 0}(r_{\rm cyl}) \left(1,8x\exp\left(-6(x^2+z^2)\right),0\right)^T}{1+16x^2\exp\left(-12(x^2+z^2)\right)}\,.
\label{eq:MagBow}
\end{equation}

Finally, we consider also the scenario, where an already fully formed
filament drifts into a magnetic field morphology with initially
parallel field lines \citep[e.g., ][]{inoue2017}:
\begin{equation}
\vec{B}(\vec{r}_{\rm cyl}) =   \frac{B_{\rm 0}(r_{\rm cyl}) \left(1,0,-5x(2-z)^2\exp(-8x^2)\right)^T}{1+25x^2(2-z)^4\exp(-16x^2)},   
\label{eq:MagFlow}
\end{equation}
where the amplitude of the disturbance depends on the z-coordinate (
see panel (f) in Fig.~\ref{fig:Models}).

As for the magnitude of the magnetic field strength we consider two
cases. First, we assume the filament to be magnetized with a constant
field strength of $B_{\rm 0}(r_{\rm cyl})=100\ \rm{\mu G}$. In the
second case we apply the familiar scaling-law
\citep[][]{Crutcher1993,Crutcher1999} of field strength with volume
density $n_{\rm gas}$:
\begin{equation}
B_{\rm 0}(r_{\rm cyl}) \propto n_{gas}^{0.6}(r_{\rm cyl}). 
\label{eq:Brad}
\end{equation}
Here, we re-scale in order to obtain a field strength at ${B_{\rm
    0}(r_{\rm cyl}=0)=100\ \rm{\mu G}}$ in the center of our filament
model.

\section{Radiative transfer (RT)}

\subsection{Molecular abundances and gas temperature}
\label{sect:MolecularAbundances}

We use the spectral synthesis code
CLOUDY\footnote{http://www.nublado.org/} $\rm v17.00$
\citep[][]{Ferland2017} to calculate the total gas temperature $T_{\rm
  g}$ and molecular abundances of the $\rm{HI}$, $\rm{OH}$, $\rm{CN}$,
and $\rm{SO}$ along the radial density profile of our filament
model. The results are shown in Fig.~\ref{fig:TempAbundance} (see also
Table~\ref{tab:ParameterMolecules}). The abundances are calculated
under the assumption of steady state. Heating and cooling are assumed
to be in local equilibrium, with the temperature and abundance
gradient set by the density gradient and the attenuated incident
radiation field and cosmic ray ionization rate. This provides heating
as well as increases the abundance of free electrons. Two 1D
calculations are performed, each for a power-law profile with the
slope $\beta$ equal to $1.6$ and $2.0$, respectively, with initial
densities. Milky Way conditions were assumed by adopting Orion nebular
metal abundances \citep{Baldwin1991}. Calculations assume a constant
Galactic cosmic ionization rate. The calculations were
stopped when an equilibrium temperature was reached.

\subsection{Grain alignment and dust heating}
\label{sect:GrainAlignment}

For the dust heating and polarization calculations we apply the code
RT code POLARIS \citep[][]{Reissl2016}. The implementation of the dust
heating follows the Monte-Carlo based method presented in
\cite{Lucy1999}. This method assumes that the dust grains exist in
thermal equilibrium with their environment:

\begin{equation}
\frac{\dot{E}_{\rm{0}}}{V}  \sum_{\rm{i}}{C_{\rm{abs,\lambda}} l_{i}} = 4 \pi \int{C_{\rm{abs,\lambda}}B_{\rm{\lambda}}(T_{\rm{dust}})d\lambda}\, ,
%\label{eq:Frad}
\end{equation}
where a photon deposits an energy of $\dot{E}_{\rm{0}}$ per unit time
in a cell of volume $V$ along its path $l_{\rm i}$ between two cell
walls. Comparing this energy content with the blackbody spectrum
$B_{\rm{\lambda}}(T_{\rm dust})$ modified by the cross section of
absorption $C_{\rm{abs,\lambda}}$ allows the derivation of the dust
temperature $T_{\rm dust}$ assuming typical Milky Way conditions.

The usual way to quantify polarized radiation is with the help of the
Stokes vector $S=(I,Q,U,V)^T$. The Stokes
parameter $I$ stands for the total intensity, whereas $Q$ and $U$
quantify the linear polarization and $V$ the circular
polarization. POLARIS solves the RT problem self-consistently in all
four Stokes parameters simultaneously
\citep[][]{Whitney2002,Reissl2016}. The polarization state of observed
radiation is then completely defined by the degree of linear
polarization, 
\begin{equation}
P_{\rm l}=\sqrt{\frac{Q^2+U^2}{I^2}}\,,
\end{equation}
the orientation angle, 
\begin{equation}
\chi=\frac{1}{2}\tan^{-1}\left(\frac{U}{Q}\right)\,,
\end{equation}
as well as the degree of circular polarization
\begin{equation}
P_{\rm c}=\frac{V}{I}\,.
\end{equation}
Note that $P_{\rm l}$ is always positive while the $P_{\rm c}$ can
also have negative values for light with circular polarization
rotating counter clockwise (depending on the convention) in direction
of the observer.

The local dust polarization within the model varies depending on the
local conditions of the model as well as the grain parameters. Hence,
another model parameter important for the polarization measurements of
filaments is the alignment efficiency of the dust grains. In contrast
to previously attempts to model dust polarization
\cite[e.g.][]{Fiege2000b}, the RT code POLARIS provides the full
spectrum of available grain alignment theories. 

The physics of grain alignment is still a field of ongoing research
\cite[see][for review]{Andersson2015}. However, the most dominant cause and
widely accepted mechanism of dust grain alignment is by means of
radiative torques (RAT) \citep[][]{Weingartner2003,Lazarian2007}. In
order for paramagnetic grains to align efficiently with the magnetic
field direction they must spin with a sufficiently large 
angular momentum $J$. Irregularly shaped dust grains can spin up by a
directed beam of radiation as well as gas collisions. The later effect
leads to an angular momentum $J_{\rm gas}$ pointing in a random
direction. Hence, one criteria for grain alignment is if the spin up
process is dominated by RATs. As it is shown in \cite{Hoang2008} a
stable alignment can only take place when the rotational angular
momentum $J_{\rm rad}$ induced by RATs is about a factor of $3 \times$
larger than the angular momentum $J_{\rm gas}$ by dust-gas
collisions. This necessary condition can be expressed as
\citep[][]{Draine1996,Draine1997,Weingartner2003}: 
\begin{equation}
\begin{split}
\left(\frac{J_{\rm{rad}}}{J_{\rm{gas}}}\right)^2 = \frac{ a_{\rm{alg}} \rho_{\rm{dust}}}{\delta 
m_{\rm{H}}} \times \qquad\qquad\qquad\qquad\qquad\qquad\\\qquad\qquad  \left(\frac{t_{\rm{gas}}}{(t_{\rm{gas}}+t_{\rm{rad}})n_{\rm{gas}}k_{B} 
T_{\rm{gas}}} \int Q_{\rm \Gamma}\lambda\gamma_{\rm{\lambda}} 
\overline{u}_{\rm{\lambda}} d\lambda \right)^2.
\end{split}
\label{eq:OmegaRatio}
\end{equation}

%%alternative presentation, but it also does not look great. 
%%\begin{equation}
%%\left(\frac{J_{\rm{rad}}}{J_{\rm{gas}}}\right)^2 = 
%%\label{eq:OmegaRatio}
%%\end{equation}
%%$$
%%= \frac{ a_{\rm{alg}}\rho_{\rm{dust}}}{\delta m_{\rm{H}}} \left(\frac{t_{\rm{gas}}}{(t_{\rm{gas}}+t_{\rm{rad}})n_{\rm{gas}}k_{B} 
%%T_{\rm{gas}}} \int Q_{\rm \Gamma}\lambda\gamma_{\rm{\lambda}} 
%%\overline{u}_{\rm{\lambda}} d\lambda \right)^2.
%%$$

Here $m_{\rm H}$ and $k_{\rm B}$ are the hydrogen mass and Boltzmann's constant, respectively. The density of the
grain material $\rho_{\rm dust}$, the geometric factor $\delta$ as
well as the grain alignment efficiency $Q_{\rm \Gamma}$ are defined by
the choice of the dust grain model as it is defined is
Sect. \ref{sect:GrainProperties}. Grain rotation is dumped down by gas
collision as well as the emission of thermal photons. These effects
are taken care by the gas dumping time $t_{\rm{gas}}$ as well as
radiative dumping time $t_{\rm{rad}}$ \citep[we refer to][for exact
  definitions]{Draine1996}. The mean energy density
$\overline{u}_{\rm{\lambda}}$ per wavelength as well as the anisotropy
factor $\gamma_{\rm{\lambda}}$ factor are calculated in a Monte-Carlo
run with POLARIS assuming typical Milky Way conditions \citep[see][for
  details]{Reissl2016}. Hence, the grain radius $a_{\rm alg}$
represents a lower threshold for effective grain
alignment. Consequently, the ratio $(J_{\rm{rad}}/J_{\rm{gas}})^2$ in
Eq. \ref{eq:OmegaRatio} amounts to a lower value in the central
regions of the filament model where the density is highest and the
radiation field is not capable of penetrating efficiently.

The magnetic field strength also plays a role in determining whether or
not grains can become aligned via RATs. Thus, a second criteria for
RAT alignment concerning the critical magnetic field strength arises
by comparing the gas dumping time with the Larmor precession time
scale \citep[see e.g.][]{Hoang2008,Lazarian2007}. However, this
criteria is always fulfilled for the particular filament model
presented in this work and is thus only mentioned for completeness.
The degree of grain alignment per grain size can be quantified with
the help of the Rayleigh reduction factor (RRF) $R(a_{\rm eff}) \in
[0:1]$ \cite[see e.g.][]{Greenberg1968,Lazarian1996} where $R(a_{\rm
  eff})=0$ means random alignment and $R(a_{\rm eff})=1$ stands for
perfect alignment (in principle, the definition of the RRF allows also
for negative values but these are irrelevant considering only RAT
alignment).

A stable dust grain alignment can either occur with the direction
angular momentum $J$ pointing parallel or anti-parallel to the
magnetic field direction. The parallel configuration comes with
perfect alignment where as at the anti-parallel one the dust grain
precesses with $J$ around $B$ \citep[][]{Roberge1999,Hoang2014}. This
case can be accounted by the factor $R_{\rm \upharpoonleft \!
  \downharpoonright}$. By introducing the ratio $f_{\rm p}$ of dust
grains aligning with the parallel configuration the RRF can be
expressed as:
\begin{equation}
R(a_{\rm eff}) = \begin{cases} f_{\rm p}+\left( 1- f_{\rm p} \right)R_{\rm \upharpoonleft \! \downharpoonright} & \mbox{if } a_{\rm eff} > a_{\rm alg} \\0 & \mbox{otherwise}   \end{cases}\, .
\label{eq:RRF}
\end{equation}
Using canonical values \citep[][]{Hoang2014,Reissl2016} it gives
${f_{\rm p}+\left( 1- f_{\rm p} \right)R_{\rm \upharpoonleft \!
    \downharpoonright}\approx0.72}$.

Two aspects of dust polarization measurements can help to deduce the
magnetic field morphology. First, any rotating dust grain aligned with
the magnetic field direction would appear spherical when observed
along a LOS parallel to the magnetic field direction. Meanwhile, an
observation perpendicular to the magnetic field lines would result in
maximal polarization (see also Fig. \ref{fig:Sketch}). For emission in the
IR and sub-mm regime, the size-averaged cross section of polarization $\Delta
C_{\rm \lambda}$ can be calculated as: 
\begin{equation}
\begin{split}
\Delta C_{\rm \lambda}  \cong \sin^2(\vartheta) \int_{a_{low}}^{a_{up}} R(a)n(a)\times \qquad\qquad\qquad\qquad \\ \qquad\qquad\qquad\qquad   \left( C_{\rm abs, \lambda, \bot}(a) - C_{\rm abs, \lambda, ||}(a) \right) da,
\end{split}
\end{equation}
where $\vartheta$ is the angle between LOS and magnetic field
direction and the cross sections of absorption $C_{\rm abs, \lambda,
  \bot}(a)$ and $C_{\rm abs, \lambda, ||}(a)$ are perpendicular and
parallel, respectively, with respect to the minor and major axis of a
spheroidal dust grain. Hence, no polarization can take place along the
LOS.

Second, due to a differential phase lag for
different polarized states, a portion of the linear
  polarization passing through the material obtains a small degree of
  of circular polarization \citep[for details we refer to][]{Martin1971,Whitney2002,Reissl2016}.
	This conversion is most
effective in case when the dust grain alignment neither parallel nor
perpendicular to the LOS. Consequently, circular dust polarization is
an indicator of non-parallel magnetic field lines along the LOS.
Formally there can be also be a contribution to circular polarization
dure to dust scattering.  However, since the scattering cross sections
are minuscule at long wavelengths we ignore this effect in this paper.

\subsection{Line of  sight (LOS) Zeeman effect}
\label{sect:Zeeman}

The Zeeman effect provides the means to observe the LOS magnetic field
strength \citep[e.g.,][]{Crutcher1993,Crutcher1999}. Certain molecular
energy levels can split into sub-levels in the presence of a magnetic
field. This gives rise to counter clockwise ($I_{\rm ccw}$) and
clockwise ($I_{\rm cw}$) circularly polarized
radiation, respectively. Hence, the Stokes parameters of intensity and circular
polarization are determined by
\begin{equation}
I = I_{\rm cw} + I_{\rm ccw} 
%\label{eq:VectorNormal}
\end{equation}
and
\begin{equation}
V = I_{\rm cw} - I_{\rm ccw} = \frac{dI}{d\nu}\Delta \nu_{\rm z} \cos(\theta).
\label{eq:Zeeman}
\end{equation}
Here, $\theta$ is defined as the angle between the LOS and the magnetic
field direction. The frequency shift caused by the Zeeman splitting is
defined as
\begin{equation}
\Delta \nu_{\rm z}=\frac{B\mu_{\rm B}}{h}\left(g'M'-g''M''\right),
\end{equation}
where $\mu_B$ is the Bohr magneton, $h$ is the Planck constant, and
$M$ and $g$ are the magnetic quantum number and Lande$\acute{e}$
factors of the lower sub-level (superscript ${'}$) and upper sub-level
(superscript ${''}$), respectively (see also
Tab. \ref{tab:ParameterMolecules}).

We perform line RT simulations with the POLARIS code
\cite[see][for details]{Brauer2017}. POLARIS can solve the line RT problem
considering the level populations of a certain molecule including
Zeeman spitting based on the physical parameter taken from
\texttt{L}eiden \texttt{A}tomic and \texttt{M}olecular
\texttt{DA}tabase
LAMDA\footnote{http://home.strw.leidenuniv.nl/$\sim$moldata/}
\citep[][]{Schoeier2005}. For calculating the level populations we
consider the conditions of local thermodynamic equilibrium (LTE), or 
alternately use the free-escape probability (FEP) implemented in POLARIS. The
later assumes that the radiation interacts with the molecule only once
and escapes then freely from the cube of the model. Both LTE as well
as FEP deliver almost identical results for our filament model. Hence,
all the results presented here are calculated with the computationally
lighter LTE condition.

\begin{table}
\centering
\setlength{\extrarowheight}{2pt}
\begin{tabular}{ l | c | c | c | c }
  Molecule & $\rm{HI}_{2-1}$ & $\rm{OH}_{3-1}$ & $\rm{CN}_{4-2}$ & $\rm{SO}_{4-3}$ \\
  \hline
  \hline
  $\nu_0\ [\rm{GHz}]$ & 1.420 & 1.665 & 133.171 & 99.30\\
  $\nu_{\rm z}\ [\rm{Hz/\mu G}]$ & 2.80 & 3.27 & -0.21 & 1.01\\
\end{tabular}
\caption{Characteristic frequency $\nu_{\rm 0}$ and Zeeman shift $\nu_{\rm z}$ for the different molecules considered in this work.}
\label{tab:ParameterMolecules}
\end{table}

The characteristic transition frequency $\nu_{\rm 0}$ between
 molecular sub-levels is broadened by Doppler
shifting. Here we take total velocity to be
${v^2=v^2_{rad}+v^2_{turb}}$ (see
Sect. \ref{sect:RadialProfiles}). Additionally, POLARIS takes the
effects resulting from natural and collisional broadening as well as magneto-optical effects,
 as presented in
\cite{Larsson2014}, into account, while performing line RT with Zeeman
splitting.

Finally, the remaining parameters are the magnetic field strength $B$
and the $\cos(\theta)$. These quantities be indirectly determined by
least square fitting the Stokes $V$ parameter in Eq. \ref{eq:Zeeman}
 resulting from the POLARIS line RT
simulations to $dI/d\nu$. In this work we are dealing with idealized
and synthetic observational conditions and the LOS magnetic field
strength
\begin{equation}
B_{||}=B\cos(\theta)
\end{equation}
can be inferred from synthetic Zeeman observations by a single
parameter fit (instrumental effects may require additional
parameters).  Note that LOS and $B$ can be either parallel or
anti-parallel because of the $\cos(\theta)$. Hence, for Zeeman
measurements the magnitude of $B_{||}$ as well as its sign can provide
valuable information about the observed projected magnetic field
morphology.  The necessary quantities for the fitting process are
listed in Tab. \ref{tab:ParameterMolecules}.

\section{Discussion}
\subsection{Dust polarization measurements}
\label{sect:ResultsDust}

We perform RT dust polarization simulations along the observer plane
(see black dashed lines in panel (a) of Fig.~\ref{fig:Models}) for the
different magnetic field morphologies in order to investigate the
emerging polarization pattern as a function of wavelength and filament
inclination.  Here we consider the {\it Herschel} \citep{Pilbratt2010}
bands
${\lambda\in [160\ \rm{\mu m},\,250\ \rm{\mu m},\,350\ \rm{\mu
    m},\,500\ \rm{\mu m}]}$,
similar to the high frequency bands of ALMA, and rotate the model by
an inclination of $i \in [0^{\circ},90^{\circ}]$ in steps of
$15^{\circ}$ around the x-axis (see Fig.~\ref{fig:Models}),
 where $i = 0^{\circ}$ is a LOS perpendicular to the symmetry
 axis of the filament.

The resulting degrees of polarization and the orientation of
polarization vectors are rather similar for the applied regime of
wavelengths. The polarization appears also to be only mildly affected 
by the power law index $\beta$ of the density profile and
the radial dependence of the magnetic field strength. Hence, we focus
in our discussion on a wavelength of $\lambda = 500\ \rm{\mu m}$, a
slope of $\beta=2.0$ and the radially dependent magnetic field case
(see Eqn.~\ref{eq:Brad}).

In Fig.~\ref{fig:PolarizationAll} we present plots of the orientation
and degree of linear polarization $P_{\rm l}$ as well as the degree of
circular polarization $P_{\rm c}$. In the upper row we show the
results for the model $'\rm{toro}'$. For an inclination of
$i=0^{\circ}$ the toroidal field exhibits a pattern of polarization
vectors that are parallel to the long-axis of the filament. Note that
this pattern is in thermal emission. Hence, the magnetic field can be
inferred along the perpendicular direction. This orientation pattern
would also be characteristic for magnetic field lines with an orientatino parallel
to the x-axis. However, the degree of linear polarization is
different. A purely parallel field would be rather constant with
decreasing radius $r$ with a minor drop close to $r=0\,\rm{pc}$. This
drop is a result of an inefficient grain alignment close to the
symmetry axis of the filament (see
Sect. \ref{sect:GrainAlignment}). In contrast to a parallel field a
toroidal field has components perpendicular to the LOS (the same as
scenario (A) in Fig. \ref{fig:Sketch}). Consequently, the degree of
linear polarization peaks toward the center. Here, the magnetic field
morphology is the dominant parameter compared to grain
alignment. Since the toroidal morphology has no crossing field lines
along the LOS we can observe no circular polarization signal at zero
inclination. For a toroidal field the overall pattern of polarization
orientation does not change between $i=0^{\circ} - 30^{\circ}$. At
this characteristic value the magnetic field lines start to cross
(see scenario (B) in Fig.~\ref{fig:Sketch}). As a consequence, linear
polarization cancels out at a certain radius and circular polarization
starts to emerge from the filament. For the purely toroidal field, the
degree of circular polarization increases with increasing inclination
angles. Both the linear as well as the circular polarization patterns
are symmetric with respect to the symmetry axis of the filament.  The
cancellation points in the polarization degree coincide with the
location where the polarization changes its direction by $90^{\circ}$,
which we term the ``flipping point'' of the orientation vectors. A
flipping point is a characteristic feature of the projected field
morphology and not a consequence of grain alignment and thus we
suggest that it may be a very useful diagnostic of the underlying field
morphology.

We present the dust polarization results of the model $'\rm{cont}'$ in
the second top row of Fig.~\ref{fig:PolarizationAll}. The amount of
$P_{\rm l}$ at $r=0\,\rm{pc}$ is not the absolute maximum of the plot
but is slightly reduced by the inefficient grain alignment close to
the center of the filament. In comparison with the toroidal field this
morphology has also no crossing field lines along the LOS and, hence,
no circular polarization for zero inclination and the polarization
patter are rather similar. In contrast to model $'\rm{toro}'$ the
model $'\rm{cont}'$ has its field lines parallel to the LOS for
$i=0^{\circ}$ and $r=0\ \rm{pc}$. As a result of this, there is no
measurable degree of linear polarization close to the symmetry axis of
the filament. As the inclination increases, the central field lines
would go from parallel to perpendicular with respect of the direction
of the LOS. Hence, the amount of $P_{\rm l}$ increases with increasing
inclination. However, the central magnetic field lines do not cross
independent of inclination. Thus, the degree of circular polarization
$P_{\rm l}$ remains at a minimum at $r=0\ \rm{pc}$. In contrast to
toroidal and helical fields the distance between flipping pints
increases with increasing inclination.

The next row in Fig. \ref{fig:PolarizationAll} shows pattern and
degrees of polarization for the model $'\rm{bow}'$. In contrast to all
models previously discussed, this model has no flipping points at
all. Indeed, the field lines do cross along several LOS with
decreasing observer plane while at the center all lines are parallel
again. Hence, two characteristic lobes are present in the degree of
linear polarization $P_{\rm l}$. However, adjacent field lines seem
never to be parallel along the LOS. Hence, $P_{\rm l}$ can never go to
zero and the polarization vectors do not flip. Furthermore, the
overall polarization pattern changes only slightly with increasing
inclination. The small drop in circular polarization $P_{\rm c}$ can be
traced back to the diminished grain alignment in the center of the
filament.

Finally, the bottom panels of Fig. \ref{fig:PolarizationAll} show the
polarization behavior the model $'\rm{flow}'$ for the RT simulations.
While models $'\rm{bow}'$ and $'\rm{flow}'$ are intended to
model completely different scenarios for how magnetic field morphologies may
be warped by a moving filament, their polarization patterns are very 
similar. Yet again, model $'\rm{flow}'$ shows no signatures of flipping
points. Whereas the pattern of polarization vectors goes from a
vertical polarization to almost diagonal for $'\rm{flow}'$ and an
increasing inclination this trend is reversed for $'\rm{bow}'$. For
model $'\rm{flow}'$ the degree of $P_{\rm c}$ shows also two
characteristic minimums comparable to those of the model
$'\rm{bow}'$. However, for model $'\rm{flow}'$ these minimums do not
arise from crossing field lines but are a result of the vertical
components of the warped field at a radius of about $r\approx
5\ \rm{pc}$. Circular polarization of model $'\rm{flow}'$ covers a
larger range concerning minimum and maximum values while the 
range and slope are similar to those of model $'\rm{bow}'$.
These results indicate that models $'\rm{flow}'$ and $'\rm{bow}'$
would be hardly to distinguishable using dust polarization
measurements alone.

Actual dust polarization measurements presented in \cite{Pattle2017} of the OMC 1 region in the Orion A filament appear to be similar to that shown in the bottom two rows of Fig. \ref{fig:PolarizationAll}. Here, \cite{Pattle2017} interpret their data as consistent with a scenario where a cylindrically symmetric field becomes distorted by means of gravitational fragmentation. However, such measurements  (in the absence of Zeeman information) do potentially allow for an alternative explanation because of the similarities between the models $'\rm{flow}'$ and $'\rm{bow}'$. As demonstrated in this paper, the polarization  a filament moving toward the observer, in the process sweeping up the homogenous magnetic field, causes the same dust polarization signature as the scenario of a presented \cite{Pattle2017}. Again, a complementary observational mission considering additional Zeeman measurements may help to to infer the actual field morphology in the OMC 1 region.  In summary, Zeeman observations will likely prove essential for  differentiating between models that generically appear similar in linear polarization alone.

Additionally, circular dust polarization can help reveal the magnetic field morphologies
  embedded in the filaments by their characteristic profiles. This was
  already demonstrated in \cite{Reissl2014} for globules. However, it
  needs to be emphasized, as above, that an amount $P_{\rm c}$ far
  below one percent will be challenging to detect circular dust polarization with real observations
  in the near future.

\subsection{Influence of the pitch angle on dust polarization}
The polarization pattern emerging from helical fields are highly
dependent on the pitch angle $\alpha$. In
Fig. \ref{fig:PolarizationHeli} we show this dependency for different
pitch angles and inclinations. The figure is structured in the same manner as
Fig. \ref{fig:PolarizationAll}. However, different rows show different
pitch angles for the helical configuration. With increasing $\alpha$
the helical field goes from toroidal (top rows) to poloidal (bottom
row). Hence the flipping points wander towards the symmetry axis
of the filament and the polarization pattern becomes almost parallel
for low inclination and high pitch angles. We note that the degree of
linear polarization is highest for high inclinations and low pitch
angles while this is the opposite for a high pitch angle.

Since this kind of field has crossing field lines along the LOS for the entire range of
inclination angles the polarization pattern starts again with a
constant pattern of orientation vectors at zero inclination. However,
compared to the toroidal field the vectors are already flipped as a
consequence of a pitch angle of $\alpha=45^{\circ}$. The most important
feature of helical fields is their asymmetry with respect to the
symmetry axis of the filament: compare panels b. and c. in
Fig.~\ref{fig:Models}. Hence, the number of crossing field lines is
no longer evenly distributed along all directions of the observer
plane. This asymmetry results in flipping points only appearing for
positive values of $r$. Consequently, the orientation of linear
polarization flips only for positive values of $r$ where as the
polarization pattern remains constant for negative $r$. The toroidal
field case the circular polarization changes sign at $r=0\ \rm{pc}$
where as the helical field has only positive values of $P_{\rm c}$ for
$r<0\ \rm{pc}$, where as for $r>0\ \rm{pc}$ shows both negative as well as positive
values.
	
\subsection{Characteristic distances of flipping points}
We note also a tight correlation between the radial distance
$\Delta r$ of flipping points and the inclination angle $i$. This
correlation is characteristic of the different applied magnetic field
morphologies, as noted above. In Fig. \ref{fig:DistanceToroCont}, we
show the radial distances of the models $'\rm{toro}'$ and
$'\rm{cont}'$.  For the model $'\rm{toro}'$ the distance between the
flipping points becomes narrower, whereas for model $'\rm{cont}'$ we
see the opposite trend. For the model $'\rm{toro}'$ the flipping pints
start to emerge at $i\approx 30^{\circ}$ while the model $'\rm{cont}'$
has flipping points even at $i = 0^{\circ}$ up to
$i\approx 80^{\circ}$. The general trends are almost independent of
wavelength $\lambda$ and the slope parameter $\beta$ for both models.

The same for the set of models $'\rm{heli}_{\alpha}'$ presented in
Fig. \ref{fig:DistanceHeli}. Here, we show the behavior of flipping
point for a helical field for different pitch angles and
inclinations. The distance of flipping points
  increases for model $'\rm{heli}_{\rm 45}'$ with increasing
  inclination $i$ and reaches a plateau for
  ${i \in[30^{\circ},60^{\circ}]}$ and reaches $\Delta r=0\ \rm{pc}$
  again at $i =90^{\circ}$ again. Observed under an inclination of
  $i = 0^{\circ}$ the flipping points begin to emerge over a range of
  pitch angles ${\alpha \in[25^{\circ},45^{\circ}]}$ where the radial
  distance goes from the maximum extent of the filament toward 
  $\Delta r=0\ \rm{pc}$.
	
We speculate that the trend between the distance of $\Delta r$
flipping points might help determine the inclination of a filament
provided that the underlying magnetic filed morphology can be well
enough constrained in the first place.

\subsection{Zeeman observations}
\label{sect:ResultsZeeman}

The results presented here are extracted from RT simulations with the
ZRAD module of POLARIS for the different magnetic field models that we
consider (see Fig.~\ref{fig:Models}).  We then fit the resulting
Stokes parameter as described in Sect.~\ref{sect:Zeeman} to create
synthetic Zeeman observations. We consider the characteristic
transitions listed in Tab. \ref{tab:ParameterMolecules} for an
inclination of $i = 0^{\circ}$ as well as the different cases of a
constant magnetic field strength of
${B_{\rm 0}(r_{\rm cyl})=100\,\rm{\mu G}}$ and a radial magnetic field
proportional to the volume density, as described in
Sect.~\ref{sect:FieldMorphologies}. Furthermore, we compare results of
a synthesized molecular abundance (see
Sect. \ref{sect:MolecularAbundances}) and a constant abundance of
$n/n_{\rm gas}=10^{-6}$.
Fig.~\ref{fig:ZeemanToro} shows the derived LOS magnetic field
strengths $B_{||}$ for the toroidal model $'\rm{toro}'$. Because
$B_{||}\propto\cos(\theta)$, the magnitude of $B_{||}$ drops towards
the center even for the case of a constant field strength (compare
also to the sketch in Fig.~\ref{fig:Sketch}, right panel). However,
some lines show an exceptional behavior with an increase of $B_{||}$
toward the center. If the magnetic field strength becomes too high in
comparison to the line width, the analysis method introduced in
Sect. \ref{sect:Zeeman} no longer applies \citep[this case is extensively
modeled and discussed in][]{Brauer2017}. The strongest effect is for OH and HI the
most considering the low gas temperature in the center of the filament
(see Fig. \ref{fig:TempAbundance}). Hence, the synthetic Zeeman
measurements of $B_{||}$ are less reliable at the center of our
filament model.

We note that this behavior is also highly dependent on the magnitude
of the of turbulent $v_{\rm turb}$ component of the gas velocity. 
A higher $v_{\rm turb}$ would cause a better tracing of the magnetic field strength. 
For a $v_{\rm turb} \gg 1000\ \rm{km/s}$ all the lines would show the decrease in $B_{||}$ close to the 
center that is characteristic of a toroidal
magnetic field morphology \citep[see e.g.][]{Brauer2017}. 
Therefore, the strong increase of a toroidal field close to the center would be 
better traced and almost no drop would be seen.

However, such a high values would be in
direct contradiction with observations
\citep[e.g.][]{Garcia2007,Arthur2016}.
Concerning the slopes, the results of the model $'\rm{heli}_{\rm 45}'$ are almost
identical with the profiles shown in Fig. \ref{fig:ZeemanToro}.  However, the magnitudes are about a factor
of $1.2$ to $2$ lower because of the poloidal component of the field in $'\rm{heli}_{\rm 45}'$ is no longer detectable.

The line profiles shown in Fig.~\ref{fig:ZeemanCont} result from the
model $'\rm{cont}'$. For this particular morphology the field lines
are parallel to the LOS at the center. Hence, we observe an increase
in the LOS magnetic field component $B_{||}$ with the maximum at
$r=0\ \rm{pc}$ for all considered parameters. Here, all Zeeman
profiles presented in Fig.~\ref{fig:ZeemanCont} follow the behavior
expected for model $'\rm{cont}'$.

In contrast to the dust polarization measurements the models $'\rm{bow}'$
and $'\rm{flow}'$ show a distinct behavior in Zeeman
observations. Because the magnetic field follows the gas flow in model
$'\rm{bow}'$ the magnetic filed is mostly warped in a direction
perpendicular to the LOS (compare panel e in Fig. \ref{fig:Models}). Hence, the derived
$B_{||}$ is up to $15$ order of magnitudes lower the the actual
magnetic field strength as shown in Fig.~\ref{fig:ZeemanBow}. This
renders it impossible to constrain the morphology by means of Zeeman
measurements. Model $'\rm{flow}'$ has field lines perpendicular to the
LOS at the outer edges of the model as well as in the center. Thus,
the derived magnetic field strength $B_{||}$ rises toward the center,
with a drop near the center itself. This characteristic shape and the
magnitude makes the model $'\rm{flow}'$ clearly distinct to the model
$'\rm{bow}'$.

As shown in Figs. \ref{fig:ZeemanToro}-\ref{fig:ZeemanFlow}, all simulated Zeeman observations share the common feature that underestimate the actual magnetic field strengths in the model by significant and sometimes very large factors.  This decrement is caused by the intensity averaging along a cord intersecting a filament with a radially declining volume density profile.  
This result has important implications for the observed Zeeman-derived field strengths and will be investigated in detail in an upcoming paper.

\subsection{Chandrasekhar-Fermi method}
In addition to the methods already discussed in this work, the Chandrasekhar-Fermi (CF, \cite{Chandrasekhar1953}) method does allow to determine the magnetic field component $B_{||}$, perpendicular to the LOS, by linking the dispersion in the velocities to the dispersion in the polarization orientations  \citep[e.g.][]{Pillai2016}. CF assumes that the magnetic field is frozen into the matter following the velocity fields. Moreover, CF requires equipartition between magnetic and turbulent energy densities.  If these assumptions hold, this method would provide the means to complement Zeeman measurements to estimate the total magnetic field strength.In this work both quantities are modeled with values close to what we know from observations. However, in our simple initial approach we do not have any dispersion in polarization angles arising from turbulence and CF breaks down within framework of our model.

More generally, the overarching issue here is the assumption of an B-fields coupling to the gas. Under which conditions the field may essentially be following the turbulent flow, such that the dispersion in the polarization angles can be interpreted in a statistical sense as a field strength \citep[][]{Crutcher2012,Planck2016A} remains an important but highly complex open question.  Hence, providing a physically well motivated model for predicting the $B_{||}$ by applying CF to synthetic dust polarization measurements and line RT is well beyond the scope 
of the current study.

\section{Summary and conclusions}
\label{sect:Summary}
In this paper, we present a simple model of a filament considering
several scenarios for warped 3D magnetic fields. We performed RT
simulations in order to identify the characteristic observables 
that may help to
distinguish between different field morphologies. Here, we used
sophisticated state-of-the-art RT simulations within the framework of
POLARIS \citep[][]{Reissl2016,Brauer2017} in order to the derive synthetic dust
polarization pattern and Zeeman LOS magnetic field measurements for
the magnetic field configurations presented in Fig.~\ref{fig:Models}. Our
results are summarized as follows:
\begin{itemize}
	\item We find that linear dust polarization is insufficient to constrain the underlying magnetic field morphology in filaments. Different morphologies are degenerate and result in similar dust polarization patterns  (see the lower two rows in Fig. \ref{fig:PolarizationAll}).
	\item As in \cite{Reissl2014}we show that circular dust polarization Pc in filaments provides a useful means with which to constrain the 3D magnetic field morphology, complementing linear polarization in a substantive way.  However, some field morphologies remain ambiguous.  Neither linear nor circular dust polarization provide direct field strengths; nevertheless, despite low fractions, observing dust circular polarization would provide important meaningful field information.  
	\item We find a low degree of circular dust polarization $P_{\rm c}$. This result requires further investigation to determine whether the expected low levels of $P_{\rm c}$ would be detectable by upcoming observing  machines. A more extensive investigating the effects of dust models, mass,  temperatures may reveal the necessary conditions for detecting the circular dust polarization signal.
	
	\item The magnetic field in filaments leaves an imprint that is detectable in the Zeeman line splitting through the line-of-sight field direction and strength.  The low temperatures and velocities make Zeeman measurements unreliable in the very center of filaments \citep[see][for details]{Brauer2017}. Nevertheless, the Zeeman parameters provide essential constraints to interpret, together with the dust polarization measurements, the 3D field configuration, such as providing direct field directions on either side of the filament.  
	
        \item Within the parameter space of our filament models we show that the projected LOS magnetic field observed by Zeeman measurements always underestimate the actual field strengths within the filament by large factors ($\times\ 2$ to more than $\times\ 10$). The implications of this finding will be investigated in an upcoming publication.
				\item Finally, we note that both dust polarization and Zeeman observations together are essential for constraining the 3D field morphology.  We suggest that an observing strategy consisting of cuts perpendicular to main filament axis will provide optimal diagnostic power necessary to  constrain 3D magnetic field morphologies.

\end{itemize}
We emphasize that the results presented here are highly idealized synthetic observations that capture the principal observational signatures of selected magnetic field morphologies without regarding for telescope or instrumental effects. Thus, careful consideration must be taken when comparing to real telescope observations.  For example, in the work presented here we have omitted various effects such as sensitivity, noise, and spatial filtering, all of which are likely to play an important role in the interpretation of e.g. ALMA measurements.  Moreover, a meaningful interpretation of any polarization measurements must account for noise, as opposed to only consider the geometric projections of ``pseudovectors''.  Thus we recommend careful interpretation the polarization orientation information obtained from real observations for two related if distinct reasons.  First, real observational effects may mimic differing field geometries, and second, the presence of 3D curved field geometries leaves an imprint on the observations which cannot be accurately interpreted under the assumption of a basically 2D field configuration.  In a future work we will address the observational issues mentioned above with the goal of generating synthetic observations combined with simulated instrumental effects. 

\section*{Acknowledgements}
S.R., E.W.P., and R.S.K. acknowledge  support  from  the  Deutsche  Forschungsgemeinschaft in the Collaborative Research Center (SFB 881) ``The Milky Way System'' (subprojects B1, B2, and B8) and in the Priority Program SPP 1573 ``Physics of the Interstellar  Medium''  (grant  numbers  KL  1358/18.1,  KL  1358/19.2).
A.M.S and D.R.G.S acknowledge funding from the \'Concurso Proyectos Internacionales de Investigaci\'on, Convocatoria 2015'' (project code PII20150171) and the BASAL Centro de Astrof\'isica y Tecnolog\'ias Afines (CATA)
PFB-06/2007.  AMS acknowledges funding through Fondecyt regular (project code 1180350). DRGS acknowledges funding through Fondecyt regular (project code 1161247) and thanks CONICYT for funding via the Pro- grama de Astronom\'ia, particularly the Fondo Quimal 2017 QUIMAL170001.

\bibliographystyle{mnras}
\bibliography{bibtex} % if your bibtex file is called example.bib

% Don't change these lines
\bsp	% typesetting comment
\label{lastpage}
\end{document}